\documentclass[12pt,revtex4]{emulateapj}

\shorttitle{Geometry, distance and reddening of the LMC}
\shortauthors{Inno et al.}

\begin{document}


\title{The panchromatic view of the Magellanic Clouds from Classical Cepheids. \\ 
 I. Distance, Reddening and Geometry of the Large Magellanic Cloud disk}  

\author{L. Inno\altaffilmark{1,2}, 
G. Bono\altaffilmark{3,2}, 
N. Matsunaga\altaffilmark{4}, 
G. Fiorentino\altaffilmark{5}, 
M.~Marconi\altaffilmark{6},
B.~Lemasle\altaffilmark{7},
R.~da Silva\altaffilmark{2,3,8},
I.~Soszy{\'n}ski\altaffilmark{9},  
A.~Udalski\altaffilmark{9},  
M.~Romaniello\altaffilmark{10,11}
and
H.-W.~Rix\altaffilmark{1}
}

\altaffiltext{1}{Max-Planck-Institut f\"ur Astronomy, 69117, Heidelberg, Germany, \email{inno@mpia.de}}
\altaffiltext{2}{INAF--OAR, via Frascati 33, Monte Porzio Catone, Rome, Italy}
\altaffiltext{3}{Dipartimento di Fisica, Universit\`a di Roma Tor Vergata, 
via della Ricerca Scientifica 1, 00133 Rome, Italy}
\altaffiltext{4}{Department of Astronomy, The University of Tokyo, 7-3-1 Hongo, Bunkyo-ku, Tokyo 113-0033, Japan}
\altaffiltext{5}{INAF--Osservatorio Astronomico di Bologna, via Ranzani 1, I-40127 Bologna, Italy;}
\altaffiltext{6}{INAF-Osservatorio Astronomico di Capodimonte, via Moiarello 16, 80131 Napoli, Italy}
\altaffiltext{7}{Anton Pannekoek Astronomical Institute, Science Park 904, P.O. Box 94249, 
1090 GE Amsterdam, The Netherlands }
\altaffiltext{8}{ASI Science Data Center, ASDC c/o ESRIN, via del Politecnico snc, 00133 Rome, Italy}
\altaffiltext{9}{Warsaw University Observatory, Al. Ujazdowskie 4, 00-478 Warszawa, Poland}
\altaffiltext{10}{European Southern Observatory, Karl-Schwarzschild-Str. 2, 85748 Garching bei Munchen, Germany}
\altaffiltext{11}{Excellence Cluster Universe, Boltzmannstr. 2, D-85748, Garching, Germany}

\begin{abstract}
We present a detailed investigation of the Large Magellanic Cloud (LMC)
disk using classical Cepheids. Our analysis is based on optical
($I$,$V$; OGLE-IV), near-infrared (NIR: $J$,$H$,$K_{\rm{S}}$)
and mid-infrared (MIR: $w1$; WISE) mean magnitudes.
By adopting new templates to estimate the NIR mean magnitudes
from single-epoch measurements, we build the currently most accurate, largest and homogeneous
multi-band dataset of LMC Cepheids. We determine Cepheid individual
distances using optical and NIR Period-Wesenheit relations (PWRs),
to measure the geometry of the LMC disk and its viewing
angles. Cepheid distances based on optical PWRs are
precise at 3\%, but accurate to 7\%, while the ones based on
NIR PWRs are more accurate (to 3\%),
but less precise (2\%--15\%), given the higher
photometric error on the observed magnitudes.
We found an inclination
 $i$=25.05~$\pm$~0.02~(stat.)~$\pm$~0.55~(syst.) deg,
and a position angle of the lines of nodes
P.A.=150.76~$\pm$~0.02~(stat.)~$\pm$~0.07~(syst.) deg.
These values agree well with estimates based
either on young (Red Supergiants) or on intermediate-age
(Asymptotic Giant Branch, Red Clump) stellar tracers, but
they significantly differ from evaluations based on old
(RR Lyrae) stellar tracers. This indicates
that young/intermediate and old stellar populations have
different spatial distributions.
Finally, by using the reddening-law fitting approach, we
provide a reddening map of the LMC disk which is
ten times more accurate and two times larger
than similar maps in the literature. We also found an LMC
true distance modulus of $\mu_{0,LMC}=18.48 \pm 0.10$
(stat. and syst.) mag, in excellent agreement with the currently most
accurate measurement \citep{pietrzynski13}.
\end{abstract}


\maketitle

\section{Introduction}

The Large and Small Magellanic Clouds (LMC and SMC) 
represent a unique example of star-forming, 
dwarf interacting galaxies in the Local Group. 
Moreover, the Magellanic Clouds (MCs) system is embedded 
in the Milky Way gravitational potential, thus their 
dynamical history strongly affects the evolution of our own Galaxy. 
Yet, we still lack a comprehensive understanding of the dynamical history 
of the complex system MCs-Milky Way.
From the theoretical point of view, two scenarios 
have emerged: the first-infall (unbound) scenario \citep{besla07,besla16},
and the multiple-passage (bound) scenario \citep{diaz11}.
In the former, the MCs have been 
interacting between each other for most of the Hubble time, 
also experiencing at least one close encounter ($\sim$ 500 Myr ago),
while they are  just past their first pericentric passage.
In the more classical bound scenario, the Milky Way potential determines 
the orbits of the MCs, that formed as independent satellites 
and only recently ($\sim$2 Gyr ago) become a binary system of galaxies
\citep[see ][for a thorough review]{donghia15}.

Even though it has proved to be challenging to distinguish
between the two scenarios on the basis of observational 
constraints, evidence is mounting that the Clouds
are now approaching the Milky Way for the first time \citep{besla16}.
In particular, detailed three-dimensional study of the LMC kinematics 
obtained from the Hubble Space Telescope shows that the relative orientation
of the velocity vectors implies 
at least one close encounter in the past 500 Myr \citep{kallivayalil13}.
This is further supported by the distribution of OB stars in the Clouds
and the Bridge \citep[a stream of neutral hydrogen that connects the MCs,][]{mathewson74}, 
which suggests a recent ($\sim$200 Myr ago) exchange of material
\citep{casettidinescu14}.
In this context, the irregular morphology of the Clouds 
is shaped by their reciprocal interaction.
Recent dynamical simulations \citep{besla16,besla12,diaz12} show
that the off-centre, warped stellar bar of the
LMC, and its one-armed spiral naturally 
arise from a direct collision with the SMC.
Thus, the observed morphology of the LMC 
can be directly related to its dynamical history.

This is the first paper of a series aimed at investigating 
 the morphology,
the kinematics and the chemical abundances of the
Large and Small Magellanic Clouds by adopting
Classical Cepheids as tracers of the young stellar populations
in these galaxies. 
In the current investigation, we focus our attention
on the LMC
geometry and three-dimensional structure,
by using Cepheids optical ($V$,$I$) and near-infrared (NIR, $J$,$H$ and $K_{\rm{S}}$)
period-luminosity (PL) and period-Wesenheit (PW) relations. 

The LMC viewing angles, the inclination $i$ and the 
position angle P.A. of the lines of nodes 
(the intersection of the galaxy plane and the sky plane),
are basics parameters that describe the directions towards 
which we observe the LMC disk.
The determination of such angles has major implications on the 
 the determination of the dynamical state of the Milky-Way--MCs system.
For instance, the uncertainty on the determinations on these angles
affects the quoted results on the LMC kinematics,
because they are needed to transform the line-of-sight velocities
and proper motions
into circular velocities, and, in turn, to determine the orbits of the stars.
The LMC viewing angle estimates available in the literature 
span a wide range of values.

It is somehow expected that different stellar tracers and methods
will provide different results because
$a)$ the old and young stellar populations in the LMC 
show different geometrical distributions  \citep{devauc72,vandermarel01a, cioni00,weinberg01},
and $b)$ these distributions are non-axisymmetric, 
so results also depend on the fraction of the galaxy covered by the adopted tracer. 

Viewing angles based on studies of Red Giants 
\citep[RG, $i= 34^{\circ}.7 \pm 6^{\circ}.2$,P.A.= 122$^{\circ}.5 \pm 8.3^{\circ}$,][]{vandermarel01a}
are consistent with the values found 
on the basis of RR Lyrae variable stars from the OGLE-III catalog 
\citep[$i = 32^{\circ}.4 \pm 4^{\circ}$, P.A. = 115$^{\circ} \pm 15^{\circ}$,][]{haschke12}.
New estimates only based on ab-type of RR Lyrae stars
\citep[$i = 22^{\circ}.25 \pm 0^{\circ}$.01, P.A.= 175$^{\circ}.22 \pm 0^{\circ}.01$,][]{deb14},
do not support previous findings based on the same tracers, 
but they agree with the values based on HI kinematics \citep{HI98}.
The quoted uncertainties and the limited precision in dating 
individual Red Clump (RC) stars do not allow us to single out whether 
old and intermediate-age stellar tracers display different spatial 
distributions
\citep[$i = 26^{\circ}.6 \pm 1^{\circ}.3$, P.A. = 148$^{\circ}.3 \pm 3.8^{\circ}$,][]{subramanian13}.
Moreover, LMC viewing angles based on stellar tracers younger than 
$\lesssim$600 Myr display conflicting values.
Using optical \citep[from MACHO,][]{allsman00} and NIR 
\citep[from DENIS,][]{epchtein98} data for $\sim$2,000 Cepheids,
\citet{nikolaev04} found a position angle of P.A.=150$^{\circ}$.2 $\pm$ 2$^{\circ}$.4
and an inclination of $i=31^{\circ}\pm1^{\circ}$. 

On the other hand, \citet{rubele12} using NIR measurements 
from the Vista Survey for the Magellanic Clouds \citep[VMC,][]{cioni11} found 
a smaller position angle, P.A.= 129$^{\circ}.2 \pm$13$^{\circ}$, 
and a smaller inclination, $i$ =26$^{\circ}.2 \pm$ 2$^{\circ}$. 
Interestingly enough, the position angle found by 
\citet{nikolaev04} agrees quite well with the value estimated 
by \citet{vandermarel14} using the kinematics of young stars 
($\lesssim$50 Myr).
More recently, \citet{jac16} used optical mean magnitudes for 
 Cepheids from the OGLE-IV Collection of Classical Cepheids \citep[CCs,][hereinafter S15]{sos2015},
including more than 4,600 LMC Cepheids. They found a smaller inclination 
 ($i$ =24$^{\circ}.2 \pm$ 0$^{\circ}.6$) and a larger position angle 
(P.A.=151$^{\circ}.4 \pm$1$^{\circ}.5$) when compared with \citet{nikolaev04}.   

The large spread of the values summarised above
shows how complex it is to estimate the LMC viewing angles, 
and how difficult it is to correctly estimate both the statistical and
the systematic error associated to the measurements.

In this paper we provide a new estimate of the LMC viewing angles, 
taking advantage of the opportunity to complement the
large sample of LMC Cepheids recently released
by the OGLE-IV survey,
with NIR observations.
In particular, we rely on single-epoch observations from the IRSF/SIRIUS \citep{kato07} 
survey and the 2MASS \citep{skrutskie06},
transformed into accurate NIR mean-magnitudes by adopting
the new NIR templates by \citet[][hereinafter I15]{inno15}.
We also provide new mid-infrared (MIR) 
mean magnitudes from light-curves collected by the ALLWISE- Multi-epoch-catalog 
\citep[Wide-field Infrared Survey Explorer,][]{cutri13}.
The complete photometric data-set is presented in Section~2. 
In Section~3 we derive Cepheid individual distances 
with an unprecedented precision (0.5\% from optical bands, 0.5--15\% from NIR bands) 
and accuracy (7\% from optical bands, 3\% from NIR bands).
We then use the Cepheid individual distances
to derive the LMC viewing angles by using
geometrical methods described in Section~4. Results are presented in Section~5,
and discussed in Section~6.
The use of multi-wavelengths magnitudes also allow us
to compute the most accurate and extended reddening map towards the LMC
disk available to date. A description of the method and the map
can be found in Section~7. Moreover, we use 
 the Cepheids' individual reddening
 to compute new period-luminosity
relations corrected for extinction.
The summary of the main results of this investigation and 
an outline of the future developments of this project 
are given in Section~8.

\section{Data sets}
\begin{figure*}[!ht]
\includegraphics[width=0.99\textwidth]{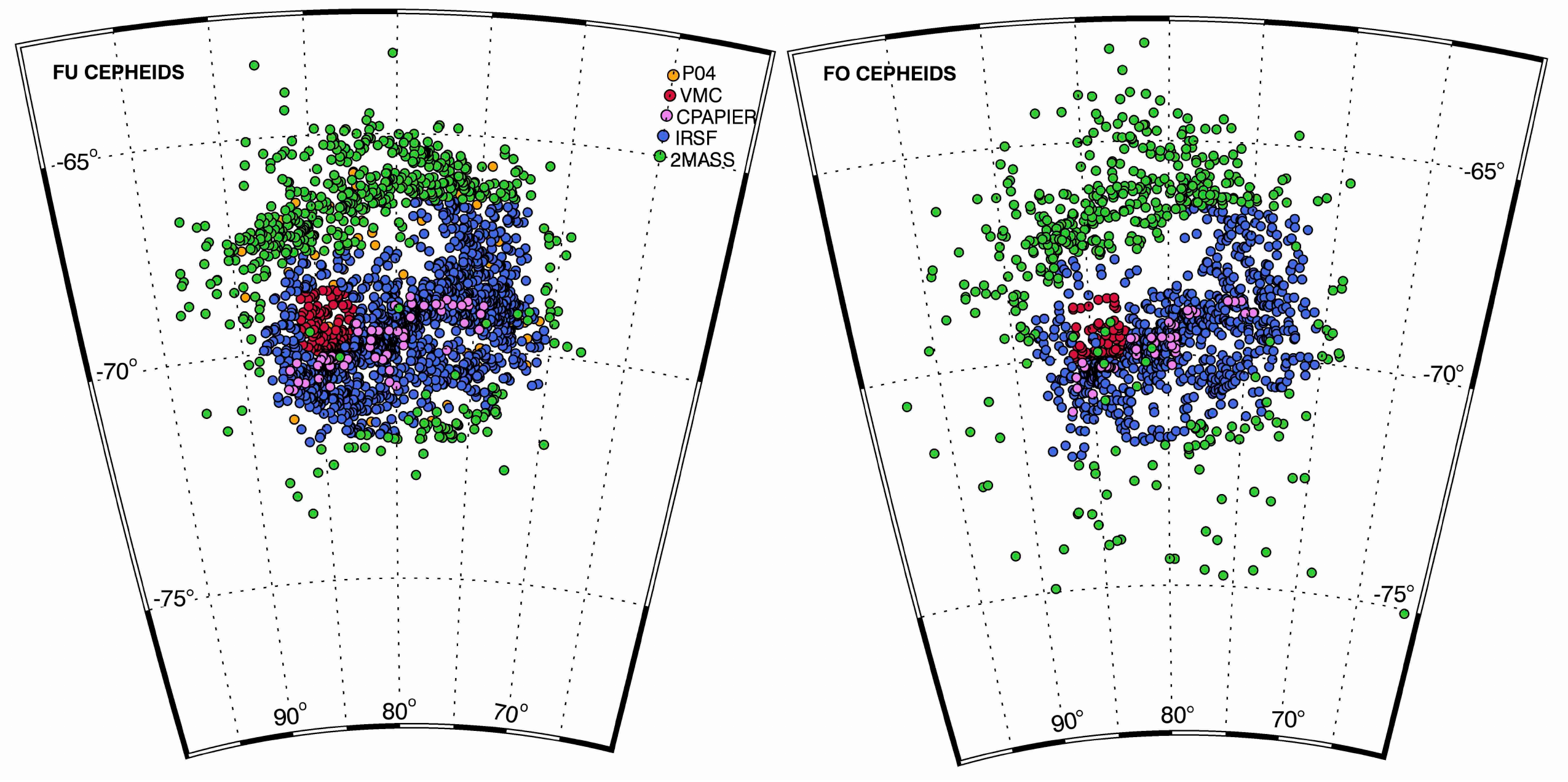}
\vspace*{0.1truecm}
\caption{Left: Sky distribution of the Fundamental Cepheids (FU) in our optical-NIR sample (Sample~A). 
Different colors indicates the different subsamples: P04 (orange dots), VMC (red dots), CPAPIER (magenta dots), IRSF (blue dots) and 2MASS (dark cyan dots).
Right: same as in the left panel but for First-Overtone (FO) Cepheids. 
A detailed description pf the number of Cepheids included in each sub-sample can be found in Table~\ref{tab1}.
Our total sample (FU$+$FO), which covers the entire LMC disk and the central bar, is the most accurate, largest and homogeneous multi-band dataset of LMC Cepheids available to date.  
}
\label{f1}
\end{figure*}

We adopted the largest dataset of NIR ($J$,$H$,$K_{\rm{S}}$) mean magnitudes 
ever collected for Cepheids in the LMC, 
that covers $\sim$95\% of the recently released new 
OGLE-IV collection of Classical Cepheids (S15). 
We define this dataset our Sample~A
and the the distribution onto the plane of the sky 
of Cepheids in this sample is shown in Figure~\ref{f1}.
Moreover, we cross-matched the LMC Cepheid catalog with 
the ALLWISE-Mep catalog, in order 
to complement our data with mid-infrared (MIR, [3.4] $\mu$m) time-series 
data. We define our Sample~B the dataset that includes only Cepheids
for which we have optical, NIR $and$ MIR mean magnitudes.  


In particular, our final catalogs includes the following sub-samples:\\

\begin{figure}[!ht]
\begin{center}
\includegraphics[width=\columnwidth]{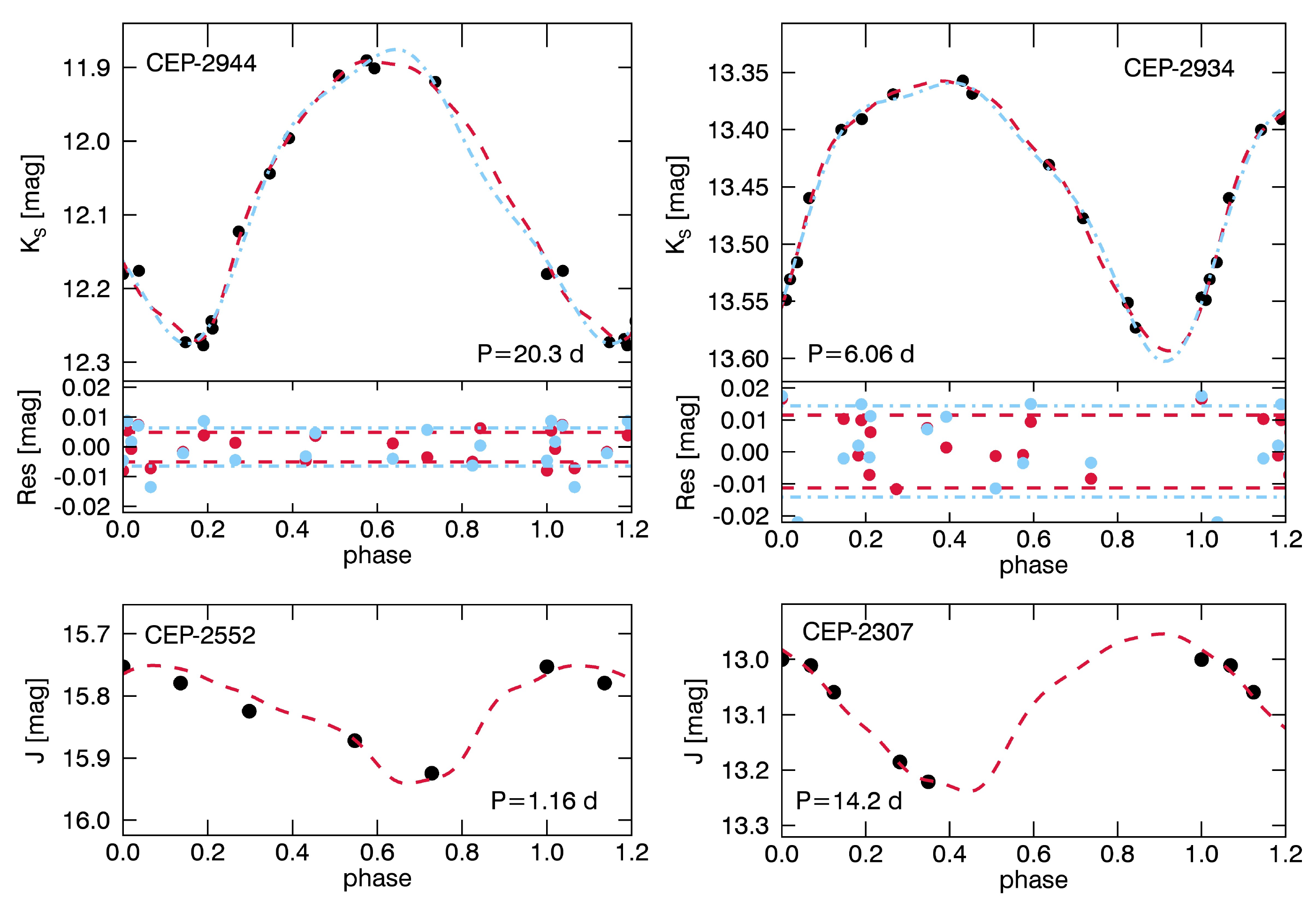}
\caption{Top --  Left: Comparison between the VMC observations in the $K_{\rm{S}}$ band (black solid dots) and two different light-curve fitting functions: 
a third order Fourier series (light blue dot-dashed line) and the template fitting by \citet[][red dashed line]{inno15} for the Cepheid OGLE-LMC-CEP-2944.
The residuals between the observed data and the fits are also shown in the lower part of the panel as light-blue dots and as red dots respectively. 
The dashed lines indicate the $rms$ of the residual, which is larger for the Fourier fitting. 
Similar results are shown in the right panel for a different Cepheid: OGLE-LMC-CEP-2934, with a shorter pulsation period. 
Bottom -- Template fitting for the five-epochs $J$-band observations of Cepheids in the VMC sample for a short period Cepheids 
(OGLE-LMC-CEP-2552, period $\sim$1~day, left panel) 
and a longer period Cepheid (OGLE-LMC-CEP-2307, period $\sim$15~days, left panel).  
The template fitting correctly recover the shape of the light curves 
for FU Cepheids over all the period range (1--100 days).
}
\label{f2}
\end{center}
\end{figure}

{\it i) multi-epochs observations for Cepheids from the VMC survey} (VMC)\\
We include the Cepheids for which the first public data release (DR1) 
of the VMC provided NIR magnitudes 
measured at five different epochs in the $J$ band and at twelve epochs 
in the $K_{\rm{S}}$ band for 157 Fundamental mode (FU)  Cepheids and 135 First Overtone (FO)
 Cepheids \citep[LMC6$\_$6 tile, see also][]{ripepi12}.
We adopted the templates by I15 and the period estimates available from 
the OGLE-IV CCs to perform a template fitting to the multi-epoch observations 
in order to determine their mean magnitudes.
We independently solved for the light-curve amplitudes, mean magnitudes and phase-lags between
the $V$ and the NIR light curves. 
Figure~\ref{f2} shows the result of the template fitting for four different Cepheids in the sample, 
in the case of the $K_{\rm{S}}$-band (top panels) and the $J$-band observations.
The top panel of Figure~\ref{f2} compares the result of the template fitting (red dashed line) 
with a third-order Fourier series fitting  (light blue dashed line)  
for two Cepheids with relatively long (P$\sim$ 20 days) and short period (P$\sim$ 6 days).
To make the difference between the two fits more clear, 
we also show the residuals in the lower part of the plots.
The $rms$ of the residuals, indicated by the dashed lines, 
show that the data have a smaller dispersion around the template fitting,
when compared to the Fourier fitting. Moreover, the bottom panel of the same figure 
shows that the template fitting is also able to properly
recover the shape of the light curve, even from few epochs. 
Without the template, only a single-sinusoide model
could be fitted to the data, but this approach fails in dealing 
with FU Cepheids, due to their asymmetric 
light curves \citep{marconi13}.

\begin{figure*}[!ht]
\begin{center}
\includegraphics[width=0.90\textwidth]{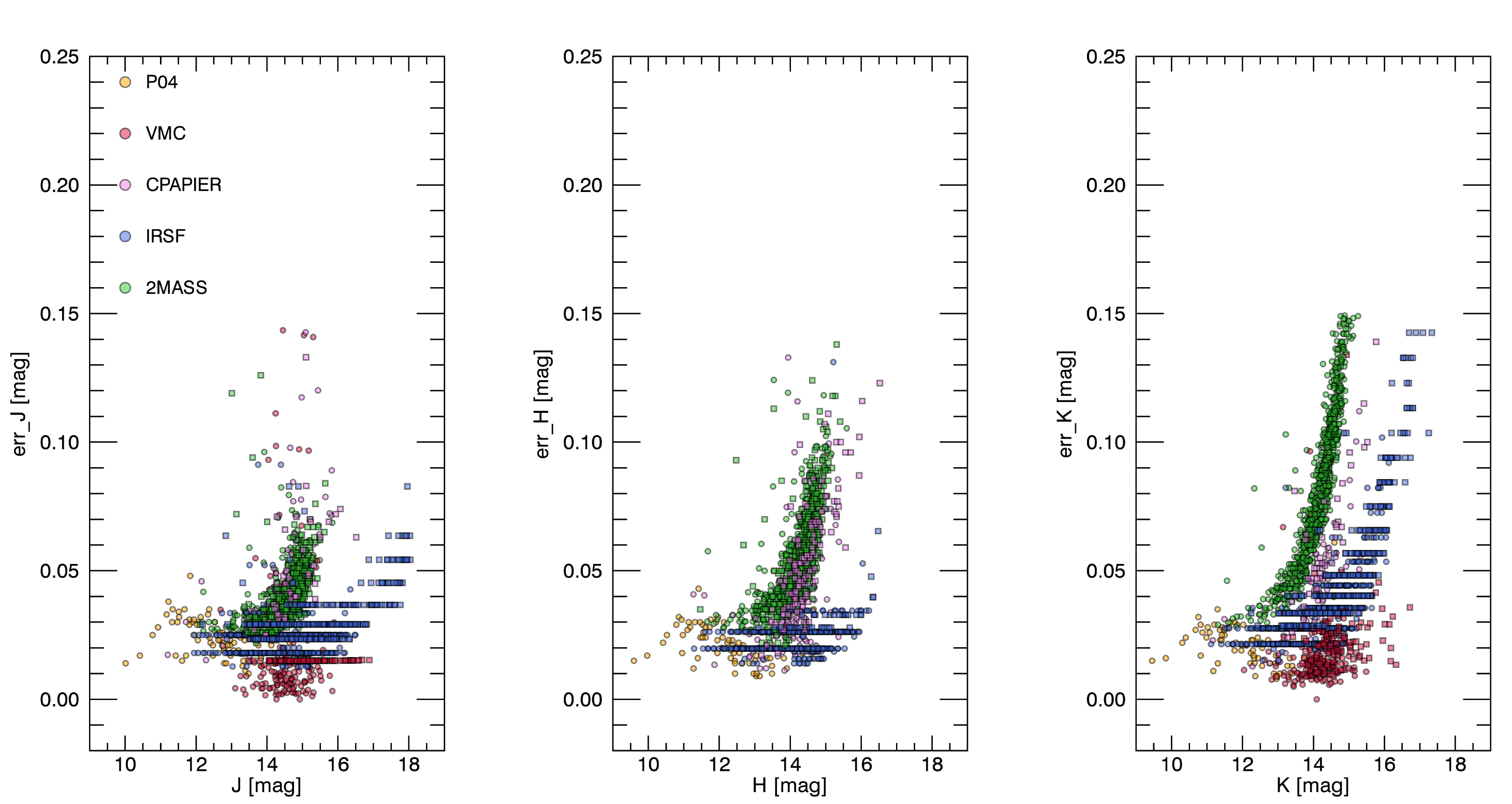}
\caption{Errors on the mean magnitudes as a function of the mean magnitudes in 
the three different NIR bands: $J$ (left), $H$ (middle) and $K_{\rm{S}}$ (right) for FU (dots) and FO (diamonds) Cepheids in our Sample~A.
The color legend is the same as in the Figure~\ref{f1}: P04 (orange dots), VMC (red dots), CPAPIER (magenta dots), IRSF (blue dots) and 2MASS (dark cyan dots).
The VMC and P04 sub-samples are characterised by the best photometric precision, while IRSF and CPAPIER data 
have  larger errors at the faint end. The data from 2MASS are characterised by the lowest photometric precision, with 
photometric errors ranging from 0.02 mag to 0.15 mag. 
}
\label{f3}
\end{center}
\end{figure*}

We also adopted template fitting for FO Cepheids 
in the $J$-band, as they are available in I15,
while we performed a third-order Fourier series fitting 
to estimate the $K_{\rm{S}}$-band mean magnitude for FO Cepheids.
The error on the mean magnitude estimates 
is given by the standard deviation of the data around the best-fit template.
The typical final uncertainty is lower than 0.01 mag.
Individual uncertainties on the mean magnitudes for the VMC sample data 
are plotted as red dots in Figure~\ref{f3}.
Note that the VMC magnitudes are already provided in the 2MASS photometric system. 
\\

{\it ii) multi-epochs observations for Cepheids from the Large Magellanic Cloud Near-Infrared Synoptic Survey} (CPAPIER)\\
 \citet{macri15} published $J$,$H$, and $K_{\rm{S}}$ light curves for 866 FU and 551 FO Cepheids 
collected by the Large Magellanic Cloud Near-Infrared Synoptic Survey, 
operated at the 1.5m CTIO telescope with the CPAPIER camera.  
The mean magnitudes and the individual photometric data, calibrated and 
transformed into the 2MASS photometric system, are publicly available.
In order to improve the accuracy on the mean magnitude determination, 
we downloaded the photometric data for all the observed Cepheids 
and performed a template fitting of the FU Cepheids light curves, 
by adopting the templates by I15 and assuming the periods from the OGLE-IV CCs. 
To perform the fitting, we adopted the same approach described above for the VMC sample.
For light-curves successfully fitted, 
the error on the mean magnitude is computed again as 
the standard deviation of the data around the best-fit template.
The mean magnitudes obtained by our fit are however very similar
to the ones obtained by \citet[][see their Table~3]{macri15}, 
In both cases, in fact, the photometric uncertainties on the individual observations, 
on average larger than 0.02 mag, is limiting the final accuracy on the mean magnitudes.  
We obtain an uncertainty on the mean magnitudes of $\pm$0.03 mag 
for  brighter ($J\approx$14 mag)  and $\pm$0.05 mag for fainter Cepheids.
In the case of the FO Cepheids, we adopted the mean magnitudes 
and uncertainties obtained by \citet{macri15}.  
Individual uncertainties on the mean magnitudes for the CPAPIER sample 
included in our final sample are plotted as magenta dots in Figure~\ref{f3}.
\\

{\it iii)  single-epoch observations for Cepheids in the IRSF sample} (IRSF)\\
 The IRSF/SIRIUS Survey provided single-epoch measurements in the 
$J$,$H$, and $K_{\rm{S}}$ bands for 1627 FU  and 1037 FO Cepheids in the LMC. 
For all these Cepheids, complete $V$-band light curves are available from the OGLE-IV CCs. 
Thus, we can apply the prescriptions and the templates by I15
to derive accurate NIR mean magnitudes 
from the single-epoch observations, by adopting the $V$-NIR amplitude ratio and the predicted $V$-NIR phase-lag.  
The final error on the mean-magnitude is also computed by following the prescriptions given in Section 6.2 of I15.
The typical uncertainty on the derived mean magnitudes is $\pm$0.02 mag 
for the brighter ($J\approx$12 mag) and $\pm$0.05 mag for the fainter Cepheids ($J\approx$ 17 mag).
\\
Light-curve templates in the $H$, and $K_{\rm{S}}$ bands for the FO Cepheids are still not available, 
due to the lack of accurate and well sampled light curves 
for short-period FO Cepheids. Thus, we can only adopt single-epoch magnitudes 
as the best approximation of the mean magnitude along the pulsation cycle. 
Thus, we can only adopt single-epoch magnitudes as the best approximation of the mean luminosity 
along the pulsation cycle. 
This approximation introduces an additional uncertainty due to the random phase effect, 
with an upper limit set by the semi-amplitude of the FO light curves in these two bands. 
From the CPAPIER data we estimate that the semi-amplitude of LMC FO Cepheids is on average lower than 0.05 mag.
Finally, we transformed the mean magnitudes into the 2MASS NIR photometric system following \citet{kato07}. 
Individual uncertainties on the mean magnitudes for the IRSF sample Cepheids are plotted as magenta dots in Figure \ref{f3}.
They typically range from  0.02 to  0.06  mag, 
when moving from brighter to fainter Cepheids.\\

{\it iv) single-epoch observations for Cepheids in the 2MASS catalog} (2MASS)\\
 We adopted 2MASS single-epoch observations 
available for all the LMC Cepheids in the OGLE-IV CCs 
that do not have NIR measurements from any of the other surveys described above. 
In order to derive the mean magnitudes of these Cepheids, 
we followed the same approach described above for the IRSF sample.
However, because the photometric precision of single-epoch 2MASS data 
is lower with respect to the IRSF data, 
the typical uncertainty on the derived mean magnitudes is larger:
$\pm$0.02 mag for the brighter ($J\approx$12 mag) and $\pm$0.10 mag for the fainter ($J\approx$ 17 mag) Cepheids. 
Nevertheless, the application of the NIR-templates allow us to
improve the accuracy on the mean magnitudes determination,
as demonstrated in Figure~\ref{f4}. From the top to the bottom,
this Figure shows the distribution of the residuals from Period-Luminosity relations
in the $J$ (top), $H$ (middle) and $K_{\rm{S}}$ (bottom) bands
for the single-epochs 2MASS data (dark green bars) and the
single-epochs + templates (lime bars). 
The distribution is fitted by a Gaussian distribution (over-plotted solid lines)
with a dispersion labeled in the left corner of each panel.
We find that the use of the templates decreases 
the residual dispersions of 5\%--10\%, thus improving 
the accuracy of the estimate mean magnitudes with
respect to the single-epoch observations.
\\

{\it v) mean magnitudes for 66 Cepheids from \citet{persson04}} (P04) \\
 We completed our NIR sample by including $J$,$H$, and $K_{\rm{S}}$ 
mean magnitudes for 66 Cepheids published by \citet[][P04]{persson04} 
and for which  $V$ and $I$ photometric data are also available in the OGLE Catalog.
The final accuracy of the NIR mean magnitudes in the P04 catalog is $\pm$0.02 mag 
for the brighter ($J \approx  12$) and $\pm$0.06 mag for the fainter 
($J \approx 14$) Cepheids. 
To transform the NIR measurements from the 
original LCO photometric system into the 2MASS photometric system,
we adopted the relations given by \citet{carpenter01}.
Individual uncertainties on the mean magnitudes 
for the P04 sample Cepheids are plotted as orange dots in Figure~\ref{f3}.
\\
%


{\it vi) mean magnitudes for 2,600 Cepheids from ALLWISE- Multi-epochs catalog (WISE)} \\
 We complemented our analysis by also including mid-Infrared (MIR) $w1$  mean magnitude 
from the ALLWISE- Multi-epochs catalog. The data are publicly available
and accessible through the IRSA web service\footnote{\url{https://irsa.ipac.caltech.edu/cgi-bin/Gator/nph-dd}}
The light curves available include
from $\sim$30 to $\sim$200 epochs for each Cepheid in the LMC.
We performed a third-order Fourier-series fit to the 
observed light curve to obtain the flux-averaged 
mean magnitude in the $w1$-band (central wavelength $\lambda_c$=3.4$\mu$m).
For light curves with scatter larger than 0.2 mag ($\sim$100 FU, $\sim$200 FU) we 
adopted the weighted mean of the measurements
as the mean magnitude.

\subsection{Compilation of the final catalogs}

For Cepheids included in different NIR sub-samples, we gave the priority to  
the data from the P04 sample and then to VMC, IRSF and CPAPIER ($H$-band) 
magnitudes. Then we also adopted CPAPIER data for Cepheids with periods 
longer than 30 days. Finally, 
2MASS observations were adopted for Cepheids with no data available from 
any other NIR catalog.

\begin{figure}[!ht]
\begin{center}
\includegraphics[width=\columnwidth]{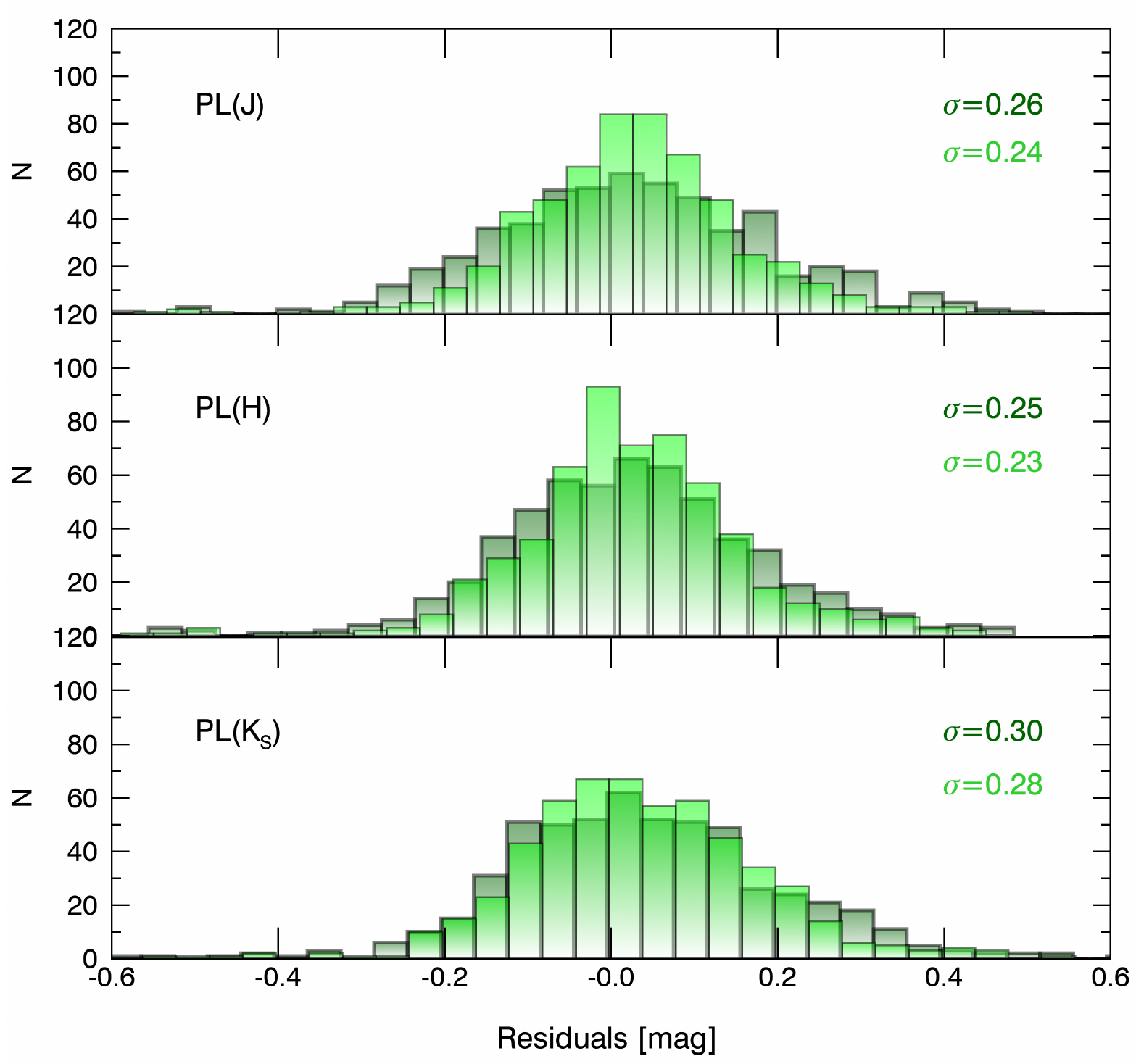}
\caption{Quantitative representation of the improvement on the determination of NIR mean magnitudes by applying NIR templates to the 2MASS single-epoch observations. \\
Top: Comparison between the residual distribution around the PL relation in the $J$-band for single-epoch (dark green) and template corrected (light green) magnitudes. In the assumption of Gaussian distribution, the standard deviations $\sigma$ of the two samples are also labeled in the top right corner. 
By applying the NIR templates, the scatter is reduced of the $\sim$8\% with respect to the use of 2MASS single-epoch magnitudes. 
Middle: The same as top but for the $H$-band.  
Bottom:The same as top but for the $K_{\rm{S}}$-band.  
}
\label{f4}
\end{center}
\end{figure}

\begin{figure}[!ht]
\includegraphics[width=\columnwidth]{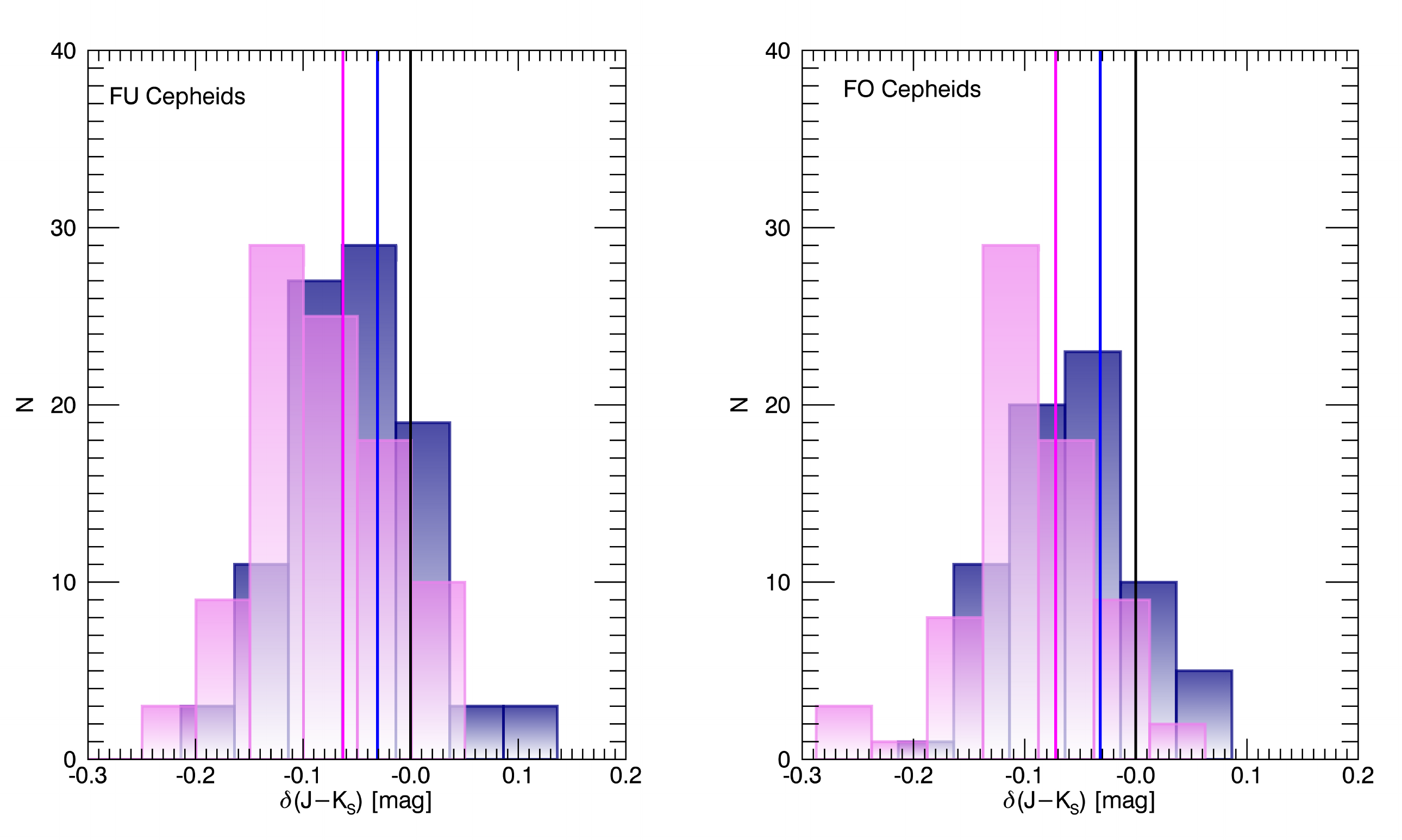}
\caption{Left: Distributions of the offsets between the mean colors of the selected reference FU Cepheids from the P04 and VMC samples 
and the ones from the CPAPIER sample (magenta bars) and from the IRSF sample (blue bars).
Right: Distribution of the offsets between the mean color of reference FO Cepheids from the VMC sample. 
and the ones from the CPAPIER sample (magenta bars) and from the IRSF sample (blue bars).
In both panels, the black solid lines indicate the reference mean color,
while the blue lines indicate the mean offset for the IRSF sample: -0.03 mag (FU,FO) and
the magenta lines indicate the mean offset for the IRSF sample: -0.06 mag (FU),-0.09 mag (FO). \\
The evidence that the mean offset of the mean colors from the CPAPIER sample is a factor of 2--3 larger
with respect to the ones in the IRSF sample demonstrates that the 
use of NIR templates together with accurate single-epoch observations provides $J$ and $K$ mean magnitudes 
that are more accurate than the ones
obtained from light curves with poor photometric accuracy.
}
\label{f5}
\end{figure}

This means that for FU Cepheids with period shorter than 30 days and FO
Cepheids, we preferred the mean magnitudes from the IRSF sample to
the ones from the CPAPIER survey.
The comparison of the 
photometric errors for the IRSF and the CPAPIER samples are 
shown in Figure~\ref{f3}. The data plotted in this Figure 
indicate that the photometric errors of the former sample is 
a factor of two smaller than the latter one for almost all the bands.
This result can be easily related to the difference in the NIR camera 
adopted by the two surveys.  In fact, both the IRSF and the CPAPIER surveys
are carried out at 1.4m telescopes (in South Africa and in Chile, respectively),
but the pixel scale of the IRSF/SIRIUS camera  (0.$"$45 pixel$^{-1}$)
is the half of the CPAPIER one (0.$"$98 pixel$^{-1}$). This means a better 
spatial resolution, and in turn, more accurate photometry in crowded stellar 
fields. 
To provide more quantitative estimates of the difference among different 
NIR datasets, Figure~\ref{f5} shows the offset in mean color $J-K_{\rm{S}}$
for Cepheids in common among different sub-samples: 23 FU Cepheids
with mean colors from the P04 sample plus 81 from the VMC sample,
for which we also have mean colors from the CPAPIER and the IRSF surveys.
We assume that the mean colors given in the P04 and VMC samples
are the \emph{reference} ones, and we compute the difference with  
the CPAPIER (magenta bars) and the IRSF sample (blue bars). 
We plot the distributions of the difference in color in Figure~\ref{f5}.
We found an average offset $\delta (J-K_{\rm{S}})$ between the 
mean colors from the CPAPIER sample and the reference data of  -0.06 mag (solid magenta line),
which is a factor of two larger than the offset from the IRSF sample (-0.03 mag, solid blue line).
However, mean magnitudes of brighter Cepheids (i.e. $\log P \gtrsim 1.4$ or $J \lesssim 12.5$ mag),
from the CPAPIER sample have a better photometric precision than the IRSF 
data, as indicated by the distribution of the magenta dots in Figure~\ref{f4}.
Thus, we decide to include CPAPIER mean magnitudes for brighter Cepheids. 
Finally, in the right panel of the same Figure, we perform a similar comparison
also for 76 FO Cepheids in the VMC sample, and we find very similar 
results: -0.09 mag for CPAPIER  and -0.03 mag for IRSF. 
As already anticipated (see, e.g., I15), we found that a single-epoch 
precise photometric measurement, together with the use of light curve 
templates, allows us to estimate individual Cepheid distances with 
an accuracy better than mean Cepheid magnitudes based on poor photometric 
quality and randomly sampled light curves. 

We end up with a sample of $\sim$4,000 Cepheids (2308 FU and 1699 FO) 
for which we obtained $I$,$V$,$J$,$H$, and $K_{\rm{S}}$ mean magnitudes (Sample~A). 
Their distribution onto the plane of the sky is shown in Figure~\ref{f1}.
For $\sim$65 \% of them we also have $w1$-band mean magnitudes.
This other sample (Sample~B) includes $\sim$2,600 Cepheids (1557 FU and 1086 FO), 
for which we have optical-NIR \emph{and} MIR mean magnitudes.
A schematic summary of the samples adopted is given in Table~\ref{tab1}. 

The new sample described here is the first optical-NIR-MIR dataset 
for Cepheids entirely covering the LMC disk, thus, allowing us to 
investigate its physical properties. In particular, we adopt the Sample~A,
which is larger, to determine the LMC disk geometry, while we adopt the
Sample~B, which includes the MIR data, to determine the reddening and the mean
distance to the LMC disk. 
The key advantage in our approach is that we are using a homogenous young stellar tracer 
for which we can also carefully quantify systematic errors.  

\section{Optical and NIR Period--Wesenheit relations}

Using the five available mean magnitudes of our Sample~A, 
and adopting the reddening law by \citet{cardelli89} with
$R_{V}$=$\frac{A(V)}{A(B)-A(V)}$=3.23 \citep{fouque07},
we can define different Wesenheit indices.
Once the reddening law has been fixed, these photometric indices 
are reddening--free pseudo-magnitudes that can be constructed using 
either two or three apparent magnitudes. The first column in Table~\ref{tab2} summarises the 
adopted Wesenheit indices.
Optical-NIR Wesenheit relations are minimally affected by uncertainties 
on the adopted reddening law, and they are also marginally affected by 
metallicity effects \citep{inno13}. 
Moreover, they are also linear over the entire period range and,
because they mimic a period-luminosity-color relations, they have 
a smaller intrinsic dispersion ($\sigma_{ID}$) caused by the width 
in temperature (color) of the Cepheid Instability Strip (IS).

Although, the intrinsic dispersion is expected to be small, 
it is not negligible. We can use up-to-date theoretical models 
for classical Cepheids to quantify it. 
The predicted IS has been computed using the non linear approach 
to stellar pulsation detailed in \citet{bono99b}
that includes a time-dependent treatment of the convection. 
We then built a synthetic population to fill this IS for the average metallicity of the LMC, i.e. Z=0.008. 
This is done by assuming a mass distribution that follows the relation M$^{-3}$ and
spans from 3 to 12 M$_{\odot}$. We then associated a period to each
synthetic star using both a pulsation and a mass--luminosity
relation given from evolutionary theory \citep[see also][]{bono99,bono99b,bono00,marconi05,fiorentino13}. 
Finally, using the updated bolometric corrections 
provided by F. Castelli\footnote{see \url{http://www.oact.inaf.it/castelli/castelli/odfnew.html}}, 
we derive the Wesenheit relations in all the desired
magnitude and color combinations.  

All the theoretical optical-NIR PW relations obtained for $\sim$1,300 synthetic stars
are listed in Table~\ref{tab2}, together with their scatter. 
The smaller is the intrinsic
dispersion (from up to down), the more accurate the determination of Cepheid
individual distances will be. An inspection of Table~\ref{tab2} shows that
NIR PW relations have smaller intrinsic dispersions when compared
to optical and optical-NIR ones.
In particular, the PW$_{JH}$ and PW$_{HJK}$ relations show the smallest dispersions, 
which are two--three times smaller than the one found around the PW$_{VI}$ relation.
Thus, Cepheid individual distances estimated on the basis of 
these relations appear to be less affected by systematic 
errors due to the intrinsic width of the instability strip.  
However, the photometric errors on observed NIR mean magnitudes 
are still significantly larger than the optical ones. 

\begin{figure*}[!ht]
\begin{center}
\includegraphics[width=0.7\textwidth]{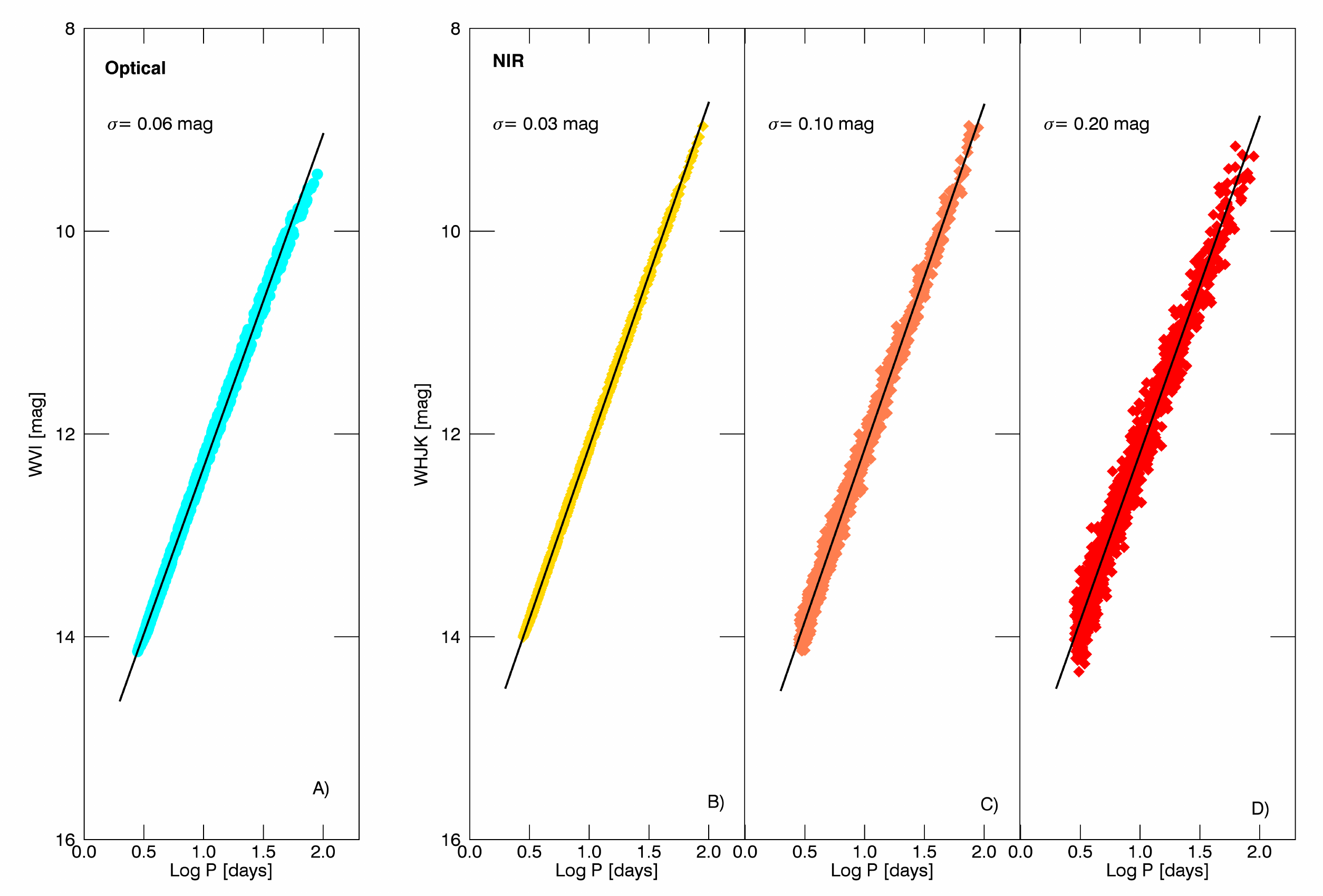}
\caption{Panel A): PW$_{VI}$ relation for theoretical models. 
The line shows the the best-fit relation, while the standard deviation around the fit is labeled in the top left corner of the Figure. 
Panel B): PW$_{HJK}$ relation for the same theoretical models. The intrinsic dispersion
due to the finite width of the Instability Strip is a factor 10 smaller respect to the one found for the optical relation 
($\sigma$=0.01 mag vs $\sigma$=0.10 mag). 
Panel C): The same relation for theoretical model but now we include photometric errors, simulated on the basis of our IRSF sub-sample. 
The dispersion around the best-fit increases to 0.12 mag because
of the photometric error. This means that the dispersion is now similar to the PW$_{VI}$  one. 
Panel D): The same as in Panel C) but the photometric error has been simulated on the basis of our 2MASS sub-sample. 
The dispersion around the best-fit increases to $\sim$0.20 mag because of the larger photometric errors. 
The comparison between the dispersions $\sigma$ in the NIR and in the optical bands shows 
that the potential of NIR PW relations for determining  accurate 
Cepheid individual distances is limited by the photometric precision of the NIR mean magnitudes available.
}
\label{f6}
\end{center}
\end{figure*}

The photometric uncertainty on the OGLE-IV mean magnitudes,
computed by adopting the standard deviation of the 7th-order 
Fourier-series fit to the $V$- and $I$-band light curves, 
is 0.007 mag for brighter Cepheids ($0.7\le \log P \le1.5$)
and 0.02 mag for fainter ones ($\log P <0.7$).
The photometric error in the $J$-band ranges from 0.005 mag 
(VMC, P04) to  0.03 mag (2MASS) for brighter Cepheids 
($0.7 \le \log P \le 1.5$) and from 0.05 to 0.15 mag (2MASS) for 
fainter ones ($\log P <0.7$).
Such larger uncertainties somehow limit the use of NIR 
photometry to determine individual Cepheid distances.
Figure~\ref{f6} shows the effect of the uncertainty on the NIR PW$_{HJK}$ relations.
The A) panel of this Figure shows the W$_{VI}$ magnitudes as a function of
the logarithmic period for our theoretical models (cyan dots) and by adopting 
an LMC distance modulus equal to 18.45 mag \citep{inno13}.
The standard deviation $\sigma$ around the best-fit relation is labeled in the top of the panel. 
This standard deviation refers \emph{only} to the $\sigma_{ID}$.
A similar plot but for the W$_{HJK}$ is shown in the B) panel of the same Figure.
In this case, the standard deviation is a half of the one around the optical PW relation.
We now simulate the photometric errors on the mean magnitudes
of the theoretical models. We adopted the photometric errors shown in 
Figure~\ref{f3} for the IRSF sample. We model the photometric error with a 
second-order polynomial function and than compute the error associated to 
each predicted magnitude from the polynomial fit and by performing a random 
extraction from a Gaussian with the same standard deviation as the 
observed one ($\sigma$=$\sim$0.3 mag).
The new relation is shown in the C) panel of Figure~\ref{f6}. 
The scatter around the relation is now due to the $\sigma_{ID}$ \emph{and} 
the photometric error on the mean magnitudes. We find that the dispersion
is now similar to the $\sigma_{ID}$ of the optical PW relations. 

Finally, we also assume the photometric errors associated to the 
2MASS sample ($\sigma$=$\sim$0.7 mag). The new PW$_{HJK}$ relation is 
shown in the D) panel of the same Figure. The standard deviation 
around the theoretical best-fit relations is now of the order of 0.2 mag, 
which is a factor three larger than the $\sigma_{ID}$ of the optical PW relations. 
Thus, the potential of NIR PW relations for accurate 
Cepheid individual distance determinations is still limited by the current 
photometric precision.
%

\subsection{Observed PW relations}

\begin{figure}[!ht]
\includegraphics[width=0.9\columnwidth]{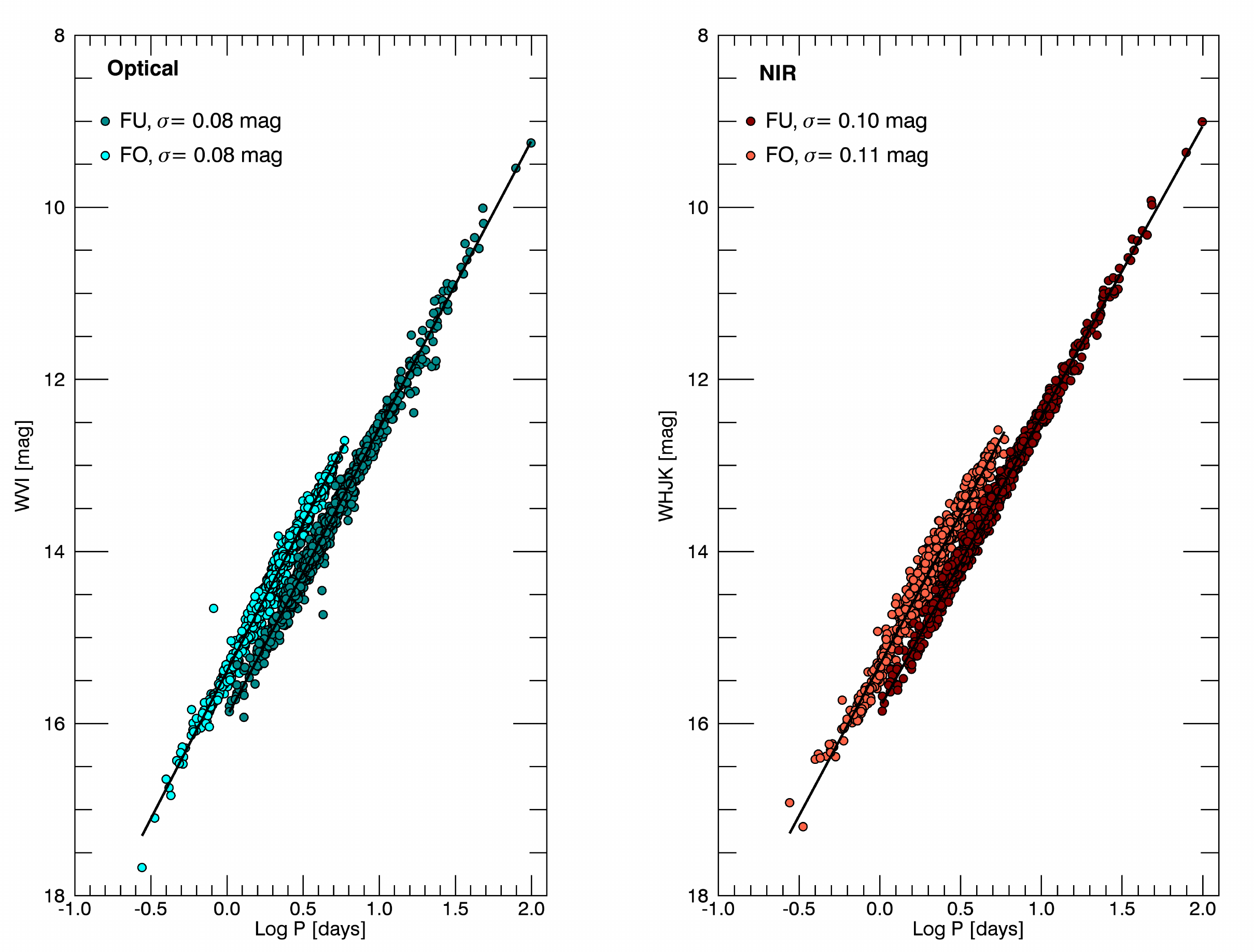}
\caption{Observed optical W$_{VI}$ (left) and NIR W$_{HJK}$ (right) PW relations  (solid lines) for FU and FO Cepheids. 
The dispersions around the best-fits are also labeled in the top.  
The residuals of the best-fit relations have been adopted to derive Cepheids' individual distances and, in turn, their 
three-dimensional distribution. The best-fit parameters of each relation are given in Table~\ref{tab3}.}
\label{f7}
\end{figure}

We derive PW relations in the form W=$a$ + $b \log P$ for all the Cepheids in our Sample~A.  
We performed an iterative
sigma-clipping \citep[biweight procedure, ][]{fabrizio11}
and a six-sigma outlier cut to perform the outlier rejection before the fitting.
The results for all the PW relations are listed in Table~\ref{tab3}, while W$_{VI}$ (left) and NIR W$_{HJK}$ (right) PW relations are 
also shown in Figure~\ref{f7}. 

Note that the current value of the slope for the PW$_{VI}$ relation 
differs at the 2.3$\sigma_b$ level with the slope found by 
\citet[$b=-3.327 \pm 0.001$ vs $b=-3.313 \pm 0.006$]{jac16}.
However, the difference
 is vanishing if we consider the dispersion around the relation, 
namely $\sigma$=0.08 mag for both the above estimates. 
The marginal difference in the slope is the consequence of 
different assumptions in dealing with outliers, namely 
3$\sigma$ \citep{jac16} vs 6$\sigma$ (ours). Using the same 
$\sigma$ clipping we find $a=15.888 \pm 0.004$ and 
$b=-3.320 \pm 0.006$, which perfectly agree with the values
found by \citet{jac16}. However, the latter cut removes 15\% 
of the Cepheids in the sample, while the former only the 3\%.
Note that the standard deviations of the PW$_{VI}$ relation is 
minimally affected by the different assumptions concerning 
the $\sigma$ clipping ($\sigma=0.08$ mag in the case of the 6$\sigma$-clipping,
and  $\sigma=0.06$ mag in the case of the 3$\sigma$-clipping).
The last column in Table~\ref{tab3} gives the standard 
deviation around the best-fit for all the PL and PW relations
derived in this investigation.

The observed standard deviation for the PW$_{VI}$ relations is, as expected, 
smaller than the ones of the PW$_{JH}$ and PW$_{HJK}$ relations, 
given the higher accuracy of the OGLE photometry.
In fact, the standard deviation decreases for all the Wesenheit indices that include optical data,
because they are affected by smaller photometric errors when compared with purely NIR Wesenheit.

By adopting all these different relations, we can obtain different distance estimates for each star, 
with associated errors that are given by the propagation of the uncertainty on the mean magnitude 
and the uncertainty on the slopes of the PW relation adopted, 
plus the systematic error given by its intrinsic dispersion, as described in the following.

\subsection{Errors on the Cepheid individual distance moduli}

The measurement error on the individual distance moduli 
obtained by adopting the PW$_{VI}$ relation is 
the sum in quadrature of the photometric
error on the Wesenheit mean magnitudes and the error on the slope $\sigma_b$,
which is anyway negligible (0.001 mag). 
This means that the
error can be propagated directly from the photometric error on the optical mean magnitudes,
and ranges from 0.01 mag (brighter Cepheids) to 0.04 mag (fainter Cepheids),
i.e. 1\%--3\% in distance.
On the other hand, theoretical predictions give us an upper limit for
the systematic error related to the ID, which is 0.06 mag, or $\sim$5\% in distance.
Summarising, while the precision of Cepheid individual distances based on the PW$_{VI}$
is better than 3\%, the accuracy is limited to $\sim$5\%. 
If we use the PW$_{HJK}$ relations, the errors on individual distance moduli
ranges from 0.03 mag ($\log P >$ 0.7 ) to 0.10 mag ($\log P <$ 0.7),
if we exclude the 2MASS sub-sample, and to 0.15 mag if we include it,
which translates to 2\%--15\% in distance.
Instead, if we choose the PW$_{JH}$ relations, 
errors are even larger (up to 20\%) as a consequence 
of the larger coefficient adopted in the Wesenheit definition ($\frac{A_H}{E(J-H)}$=1.630)
respect to the one in the W$_{HJK}$ ($\frac{A_K}{E(J-H)}$=1.046).
Moreover, \citet{inno13} found that the PW$_{JH}$ relation 
are more affected by uncertainties on the slope of the reddening law. 
Thus, we adopted the NIR PW$_{HJK}$ relation, 
which is minimally affected by such uncertainty.
In fact, the uncertainty on the assumed
reddening law also contributes to the systematics. 

Recently, \citet{demarchi16} found that in the 30 Doradus star forming region, 
the reddening law changes and in particular the total-to-selective extinction is $R_{V}=4.5$, 
thus larger than the one adopted here, i.e. $R_{V}=3.23$.
However, if we compute the coefficient of the W$_{HJK}$ 
corresponding to $R_{V}=4.5$, we find a discrepancy
lower than 1\% (1.041 vs 1.046), which minimally affects our results, 
while for the optical bands this discrepancy is of the order of 20\% (1.70 vs 1.55).
If we assume $R_{V}=4$, we find a slope of the PW$_{VI}$ relation which is steeper of 0.04 mag 
with respect to the one listed in Table~\ref{tab3}, 
while the slope of the PW$_{HJK}$ relation only changes of 0.01 mag.
Thus, the systematic error due to the uncertainty on the total-to-selective absorption ratio
is at 2\% level for the optical and at 0.5\% level for the NIR.
Moreover, the theoretical predictions in Table~\ref{tab3} indicate
that the ID for the  PW$_{HJK}$ relation is  $\lesssim$0.03 mag, 
which corresponds to an accuracy better then 2\% on the individual
distance estimates.
Thus, the precision of Cepheid individual distances based on the PW$_{HJK}$
highly depend on the sub-sample adopted, and it ranges from 2\% to 10\%,
when excluding the 2MASS sub-sample, and to 15\% when including it,
while the systematic effects are lower than 2\%. 
Concluding, Cepheids distances derived on the basis of optical relations 
are affected by significant systematics ($\sim$7\%), 
while distances derived on the basis of NIR relations 
are mostly limited by measurement errors (2\%--15\%). 
%

\begin{figure*}[!ht]
\begin{center}
\includegraphics[width=0.7\textwidth,height=6.3cm]{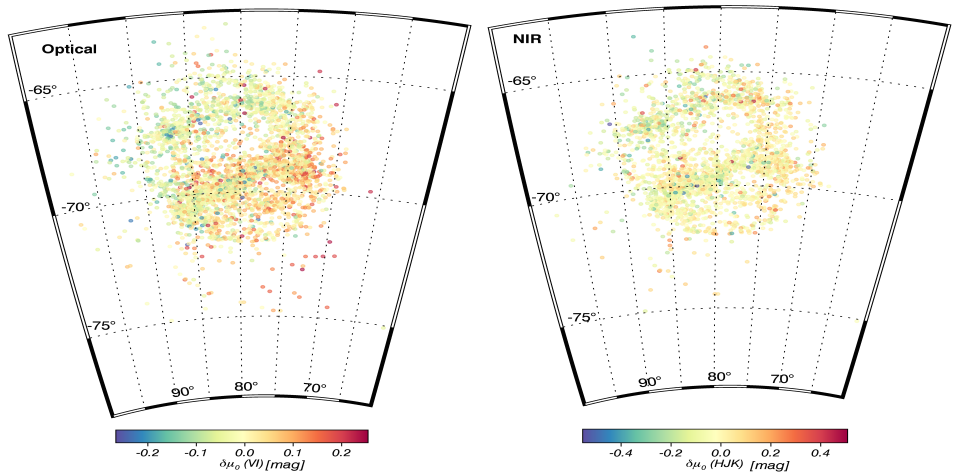}
\caption{Sky distribution of the Cepheids in the Sample~A (FU$+$FO) color coded by their relative distance moduli (mag)
obtained by adopting Equations~\ref{eq0} for the optical PW$_{VI}$ relation (left) and the NIR PW$_{HJK}$ relation (right).  
Both the distributions show similar features, with the eastern part of the LMC  closer to us  (negative relative distance moduli) respect to the western region (positive relative distance moduli).  The relative distance moduli shown here have been transformed into absolute distances according to Equation~\ref{eq1} and used to derive the viewing angles of the LMC plane. 
 }
\label{fn}
\end{center}
\end{figure*}
%

\section{Distance Measurements}

We adopted the PW$_{VI}$ and PW$_{HJK}$ relations listed in Table~\ref{tab3} 
for the FU and FO Cepheids in order to estimate the distances to all the Cepheids in our 
Sample~A. 

The individual relative distance moduli have been estimated 
by calculating the differences:
\begin{eqnarray}
\label{eq0}
\delta\mu_{0,i,VI}=W_{i,VI}-(a_{VI}+b_{VI} \log P_{i}), \\
\delta\mu_{0,i,HJK}=W_{i,HJK}-(a_{HJK}+b_{HJK} \log P_{i}), 
\end{eqnarray}
where $a$ and $b$ are the coefficients in Table~\ref{tab3} for the corresponding PW relations and $W_i$ is
the mean Wesenheit magnitude for the $i$-th star in the given bands.
Figure~\ref{fn} shows the projection onto the plane of the sky of the Cepheids in our sample,
color coded by their individual relative distance moduli obtained by adopting the optical
PW$_{VI}$ relation (left panel) and the NIR PW$_{HJK}$ relation (right panel). 
The color coding clearly shows that that the eastern parts of the LMC bar and northern
arm are closer to us (negative distance moduli) with respect to the western regions (positive distance moduli), thus
indicating that the LMC is not seen face-on, but it is inclined 
respect to the plane of the sky \citep[see also][]{weinberg01, vandermarel01a, vandermarel14,jac16}. 
In order to measure such viewing angles, we first need to 
convert the relative distance moduli given by Equation~\ref{eq0} 
into individual absolute distances (kpc).
We adopted the standard formula
\begin{eqnarray}
\label{eq01}
D_i=10^{[0.2\times (\delta\mu_{0,i}+\mu_{0,LMC})-2])},
\end{eqnarray}
where $\mu_{0,LMC}$=18.483 mag is the mean distance modulus to the LMC
\citep[][hereinafter P13]{pietrzynski13}, corresponding to the distance $D_0 = 49.97$ kpc. 
Thus, we use the individual distances $D_i$ to move into 
the cartesian reference system introduced by \citet{weinberg01} and \citet{nikolaev04}.
This new reference system ($x$,$y$,$z$) has its origin at the centre of the galaxy, 
defined by the position ($\alpha$,$\delta$,$D$)$\equiv$($\alpha_0$,$\delta_0$,$D_0$). 
The $z$-axis is pointed towards the observer, the $x$-axis is anti-parallel to the $\alpha$-axis 
and the $y$-axis is parallel to the $\delta$-axis. 
The ($x_i$,$y_i$,$z_i$) coordinates for each Cepheid are then obtained using the transformation equations
\begin{eqnarray}
\label{eq1}
x_i=-D_i \sin(\alpha_i-\alpha_0) \cos \delta_i, \ \ \  \nonumber \\
y_i=D_i \sin\delta_i \cos\delta_0 -D \sin\delta_0 \cos(\alpha-\alpha_0) \cos \delta_i, \ \ \   \\
z_i= D_0 -D_i\sin\delta _i\sin\delta_0 -D_i \cos\delta_0\cos\delta _i\cos(\alpha_i-\alpha_0)\ \ \ \nonumber. 
\end{eqnarray}
Because of its non-axis-symmetric shape, the LMC disk
does not have a very well defined center. 
Thus, the definition of ($\alpha_0$,$\delta_0$,$D_0$)
is somewhat arbitrary and \citet{vandermarel01a} showed that 
it does not affect the results.
We estimated the center of the Cepheid 
distribution by computing the Center of Mass 
as follows:
$$CM_u = \frac{\Sigma_i w_i  u_i}{\Sigma_i w_i}, $$
where $u_i$ are the coordinates $\alpha$ and $\delta$ of the $i$-th star 
and the weights $w_i$ is given by its inverse distance in the 2D-space ($\alpha$,$\delta$).  
To compare our results with similar findings from other authors, we also adopted four 
different locations for the center of the distribution available in the literature.
Table~\ref{tab4} lists all the adopted values together with their ID.
In particular, we adopted the center of the rotation map of the HI estimated by \citet[][CII$_{LMC}$]{HI98}, 
the center of the visual-band isophotes estimated by \citet[][CIII$_{LMC}$]{devauc72} 
and the geometrical center estimated by \citet[][CIV$_{LMC}$]{nikolaev04} from 2MASS data for LMC Cepheids.
\begin{figure*}[!ht]
\begin{center}
\includegraphics[width=0.7\textwidth]{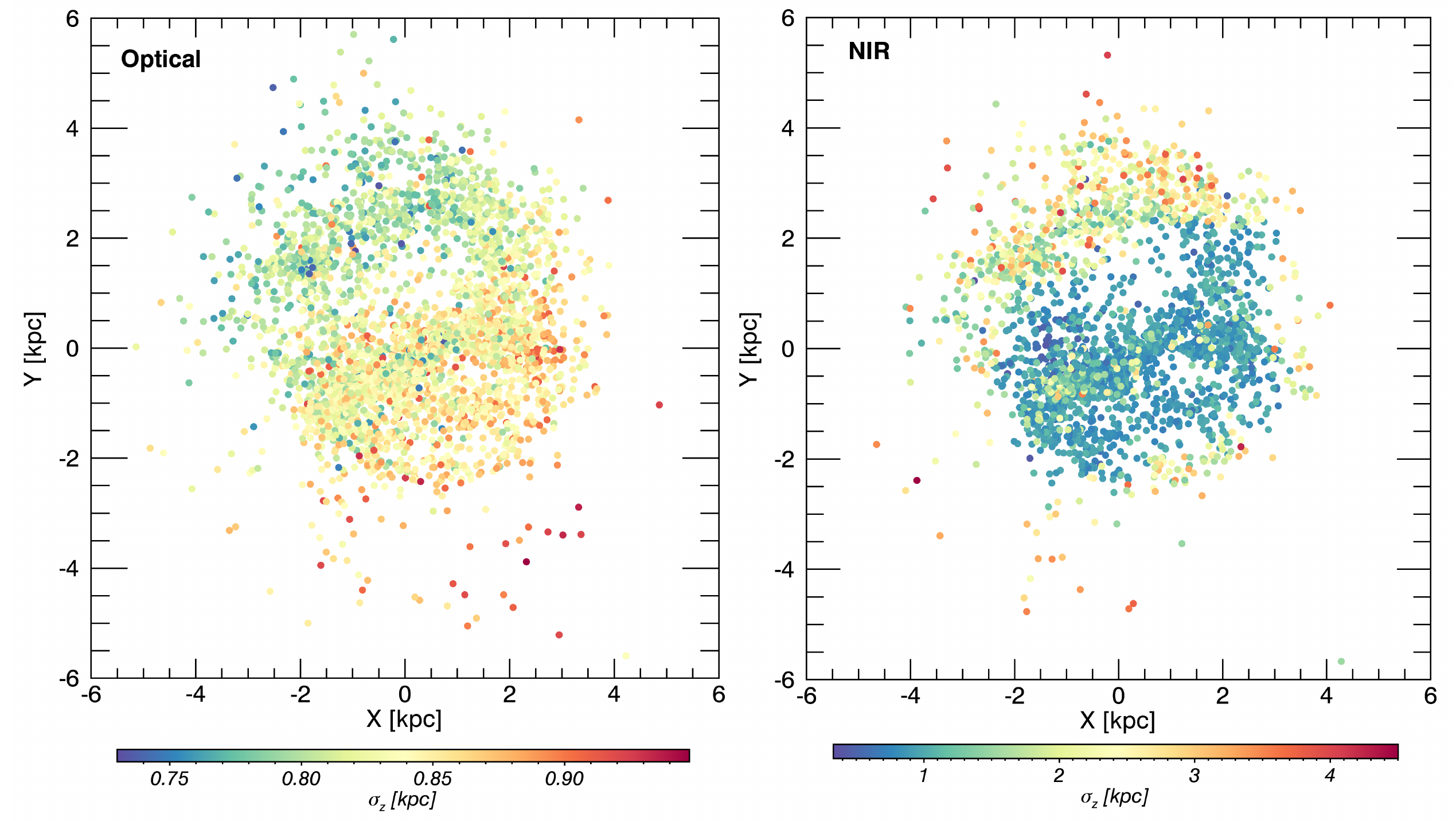}
\caption{Comparison between the measurement errors on $z$ 
in the case of distances derived from the optical PW$_{VI}$ relation (left) and
from the NIR PW$_{HJK}$ relation (right).  
The error on $z$ is obtained by propagating the uncertainty on $D_0$  and the ones on the individual distance moduli $D_i$, which are essentially given by the error associated to the mean magnitudes in the different bands. 
The errors on $z$ are similar for optical and NIR individual distances $only$ in the case of our most accurate sub-samples ,i.e. P04, VMC, IRSF and CPAPIER, while they are up to four times larger for the 2MASS sample. 
The low photometric accuracy of the 2MASS sub-sample is the main culprit of the limited accuracy of our results 
based on NIR data.}
\label{f7b}
\end{center}
\end{figure*}

\section{LMC viewing angles}

Once we have the $x_i$,$y_i$, and $z_i$  for each star in the cartesian system, 
we derive the orientation, i.e., inclination $i$ and position angle P.A, of the 
LMC disk, by fitting a plane solution of the form:
\begin{equation}
\label{eq3}
z=Ax + By+ C; \ \ \   
\end{equation}

To estimate the best-fitting plane, we performed a least-squares method 
where $x$ and $y$ has been considered the independent variables.
In fact, from Equation~\ref{eq1} follows that the error on $z_{i}$ is larger the errors
on $x_{i}$ and $y_{i}$.  
The dominant term in the error budget is given by the uncertainty on the distance $D_i$, 
since the positions on the sky ($\alpha$,$\delta$) are known with a precision better than 0$"$.2 \citep{jac16}
from the OGLE catalog.
In particular, $$\sigma_{x,y} \propto \frac{\sigma_D}{D} (x,y); $$
while the error on $z_i$ also accounts for the uncertainty on $D_0$:
$$ \sigma_{z} \propto \sqrt{\frac{\sigma_D}{D}^2 z^2+ \frac{\sigma_{D_0}}{D_0}^2}.$$
Figure~\ref{f7b} shows a comparison between the errors $\sigma_{z}$ for distances
obtained from the optical PW relations (left) and for the ones based on the NIR PW relations (right).
The Cepheids are plotted in the $x$,$y$ plane and color-coded by their error $\sigma_{z}$ on $z$, 
which ranges from $\sim$0.5 kpc (optical, NIR) to $\lesssim$1 kpc in the case of the optical data,
and to $\lesssim$4.5 kpc in the case of the NIR data. However, we find an error $\sigma_{z} >$ 1 kpc 
only for the Cepheids belonging to the 2MASS sub-sample. Thus, the limited photometric
accuracy on the 2MASS data is the main culprit of the limited accuracy of our results based on the NIR distances.

\begin{figure*}[!ht]
\begin{center}
\includegraphics[width=0.90\textwidth]{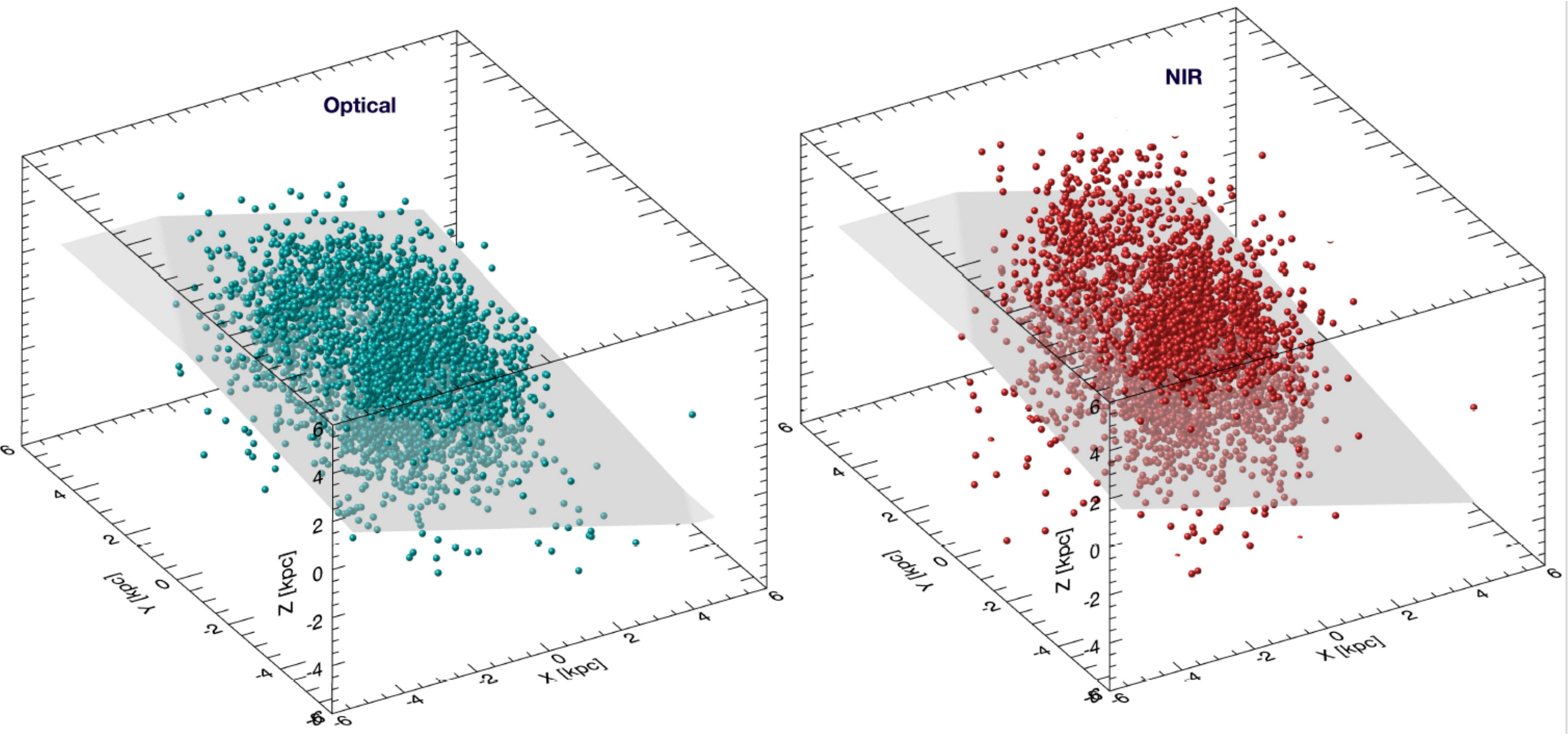}
\caption{Three-dimensional distribution of the LMC Cepheids in our Sample~A.  
Distances have been determined by adopting the optical PW$_{VI}$ relation
(left) and the NIR PW$_{HJK}$ relation (right).The grey shadowed area show the best-fit plane from which we can derive the LMC viewing angles.
A qualitative comparison between the two panels shows that the three-dimensional distribution of the LMC Cepheids as mapped by the optical data
is very similar to the one mapped by the NIR ones. 
However, the larger scatter in $z$ observed in the northern spiral arm for the right panel with respect to the left panel is a consequence 
of the larger photometric error of the NIR data from the 2MASS sub-sample respect to the optical data.  
}
\label{f8}
\end{center}
\end{figure*}

Finally, note that parameter $C$ in Equation~\ref{eq3} is introduced to remove any possible bias 
due to arbitrariness in the definition of the central position. 
The constant $C$ would be zero if the origin of our coordinate system 
($\alpha_0$,$\delta_0$,$D_0$) corresponds to the center of the LMC disk plane. 
We found $C \lesssim 10^{-3}$ kpc for our CI, 
and less than 1 kpc in the case of the adopted center: CII, CIII  CIV, and CV. 
This indicates a negligible discrepancy between these positions 
and the center of the population traced by the Cepheids. 
We can also estimate the error associated to our best-plane solution, 
by calculating the standard deviation, and we found $\sigma_{zfit}=1.7$ kpc. 
Figure~\ref{f8} shows the three-dimensional distribution of the Cepheids in the $x$,$y$,$z$ space
in the case of distances determined by adopting the optical
(left) and NIR (right) PW relations. The best-fitting planes are also shown
as shaded surfaces. 

From the coefficients $A$,$B$ and $C$ in Equation~\ref{eq3} 
we derive the position angle P.A. and the inclination $i$ of the disk
\begin{eqnarray}
P.A.=\arctan \left( -\frac{A}{B}\right) + sign (B) \frac{\pi}{2},\nonumber  \\ 
i= \arccos \left( \frac{1}{\sqrt{A^2+B^2+1}}\right).
\label{eq: ang}
\end{eqnarray}
The ensuing errors on the above angles can also be determined
by propagating the errors on the best-fit parameters 
as discussed in the following section. 
The values we found for the LMC are listed in Table~\ref{tab5},
together with a list of literature values for comparison.

\subsection{Error budget on the viewing angles}

By assuming that $C$ is a constant while the coefficients $A$ and $B$ 
are independent, we can use the standard error propagation formula,
i.e., 
\begin{eqnarray}
s_f=\sqrt{\left(\frac{\partial f}{\partial A}\right)^2 s_A^2 +\left(\frac{\partial f}{\partial B}\right)^2 s_B^2}
\end{eqnarray}
where $f$ is one of the functions defined by Eq.~\ref{eq: ang}.
Once we compute the derivative, we find: 
\begin{eqnarray}
s_{P.A.}=\frac{1}{A^2+B^2}\sqrt{B^2 s_A^2+A^2 s_B^2} , 
\label{eq: er1}
\end{eqnarray}
and 
\begin{eqnarray}
s_{i}= \frac{1}{A^2+B^2+1}\frac{1}{\sqrt{A^2+B^2}}\sqrt{A^2 s_A^2+B^2 s_B^2}
\label{eq: er2}
\end{eqnarray}
The coefficients $A$ and $B$ are equal to $A=-0.394 \pm 0.009$ and $B=  0.223\pm0.009$
when adopting the optical PW
relation, and to $A= -0.419\pm 0.011$ and $B=  0.234\pm 0.011$ when using the NIR PW
relation. 
By propagating the errors according to Equations~\ref{eq: er1}~and~~\ref{eq: er2}, 
we found $s_{P.A.}=0^{\circ}.02$ and $s_{i}=0^{\circ}.01$ 
in the case of viewing angles determined on the basis of the optical data,
and  $s_{P.A.}=0^{\circ}.02$ and $s_{i}=0^{\circ}.02$ in the case of the NIR data.
The errors associated on the coefficients are computed according to the least-square
method we adopted for the fit. We used the IDL package 
MPFIT\footnote{http://cow.physics.wisc.edu/~craigm/idl/idl.html} to perform
the fit,  which provides the formal 1-$\sigma$ error for each parameter, computed
from the covariance matrix, where individual measurements are weighted with
the inverse of the associated error.
The errors on $x_i$,$y_i$ and $z_i$ for each star have been computed as
described in the previous section.

\section{Comparison with previous estimates} 

Data listed in Table~\ref{tab5} display several interesting features worth being discussed in 
more detail. 

\begin{figure*}[!ht]
\begin{center}
\includegraphics[width=0.75\textwidth]{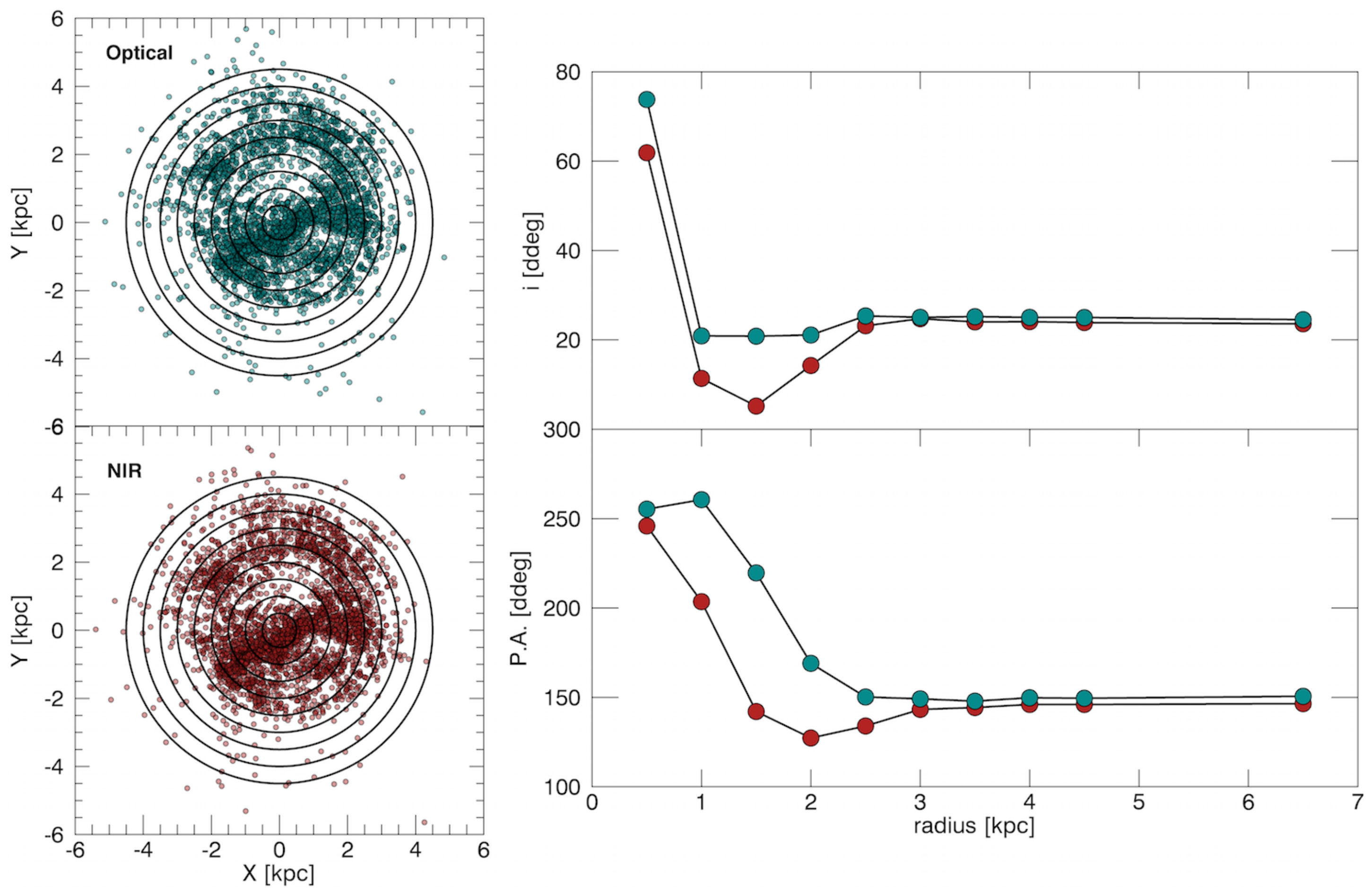}
\caption{Variation of the viewing angles of the LMC disk as a
function of the radial distance (in kpc). 
Left: The LMC disk is divided into concentric annuli with different radii, 
starting from 0.5 to 6.5 kpc with a steady increase of 0.5 kpc. The annuli 
are  over-plotted on the Cepheids' spatial distribution
in the $x$,$y$ plane based on the optical (top) and NIR (bottom) PW relations. 
Right: The upper panel shows the variation of inclination as a function of the 
radial distance, while the lower panel shows the variation of the P.A., for 
distances based either on optical (dark cyan) or on NIR (dark red) PW relations.
}
\label{fr1}
\end{center}
\end{figure*}

{\em i) Internal consistency}--The current estimates of the P.A. (column 2) and of the 
inclination (column 3) are, within the errors, minimally affected by the adopted 
center. Moreover and even more importantly, we provided independent estimates using 
optical and NIR mean magnitudes. The two data sets are affected by different measurements 
and systematic errors, as already discussed in Section~3. 
Therefore, we can perform an average to obtain our best estimate and adopt the spread 
in the values as a solid estimates of the systematic error: 
$i= 25^{\circ}.05 \pm 0^{\circ}.02$ (statistical)~$\pm 0^{\circ}.55$ (systematic) and
P.A. =150$^{\circ}.76 \pm 0.^{\circ}$20(statistical)~$\pm 0.^{\circ}07$(systematic).   
We also compared the LMC viewing angles obtained
by adopting independently either FU or FO Cepheids and we found that  
the P.A. differ of  $\sim0^{\circ}.16$ (optical) and $\sim0.^{\circ}48$ (NIR),
while inclinations differ of $\sim0^{\circ}$.4 (optical)  and $\sim3^{\circ}.5$ (NIR).
However, the values based on FU Cepheids agree within $1\sigma$ with the values 
based on the the entire sample, and with the adopted best value. 
The viewing angles based on FO Cepheids differ of about at 4$\sigma$ level ($\lesssim0.^{\circ}1$) 
with the solution based on the entire sample and with the adopted best value. 
The above difference between FU and FO Cepheids appears the consequence that FO 
Cepheids account for less than 40\% of the entire sample. The spatial distribution 
of FU and FO Cepheids shows also some diversity, with the FO Cepheids more extended
in the outer disk when compared with the FU ones (see, e.g., Figure~\ref{f1}).
To quantify the quoted variations we also investigated the change in viewing 
angles as a function of the distance from the center of the LMC. The results 
are shown in Figure~\ref{fr1}, where the variation in inclination and P.A.
are plotted as a function of the radius, in the case of distances
based on the optical (dark cyan dots) and on the NIR (dark red dots) data.
In the left panel of this Figure, the distribution in the plane $x$,$y$ of the LMC
Cepheids is shown. We defined circular regions in this plane, with radius
from 0.5 to 6.5 kpc with a step of 0.5 kpc, and considered
all the Cepheids included inside such regions to determine
the viewing angles. 
The values found at different radii are plotted in the 
left panel and show that values based on Cepheids located inside a radius
of $\sim$3 kpc from the center differ significantly from the
values based on larger area. This is not surprising, since 
the central regions are dominated by the bar and have a complex geometry.
Moreover, the discrepancy between viewing angles based on optical and NIR data
in these regions might be related to a possible change of the reddening law in the
more extincted regions of the LMC bar \citep{demarchi16}.
On the other hand, values based on the outer regions are 
in excellent agreement with each other, and there is no solid evidence 
of a change of the viewing angles with radial distance.


{\em ii) External consistency}--The current estimates of LMC viewing angles agree well 
with similar estimates based on classical Cepheids available in the literature. 
The excellent agreement between our values and the values found by  \cite{jac16}, on the basis 
of the same sample from the OGLE-IV CCs 
demonstrates once again the robustness and precision of the approach adopted. 
The position angle derived by \citet{nikolaev04} from independent datasets 
agrees also very well with our best 
estimate, while their inclination is larger. 
The difference in inclination is likely due to the improved accuracy on
mean NIR magnitude  from single epoch measurements (see Section~2 and their Section~3.2) and
also to the sample (our sample is a factor 2 larger, see also Section~8).   

The comparison with the estimates provided by P04 indicates a good agreement 
for the inclination, but a 2$\sigma$ difference in the PA.
However, their sample is again significantly smaller (92 vs $\sim$ 3,700) and biased towards  brightest stars.
Similar arguments apply to the estimate provided by \citet{haschke12}, since they only adopted 
optical mean magnitudes provided by OGLE-III, which was limited to a smaller area close to the LMC bar.

{\em iii) Age consistency}--Recent estimates of the LMC viewing angles provided by 
\citet{vandermarel14} using Red Super Giants (RSGs) agree quite well with current estimates. 
This is an interesting findings for a twofold reason: $a)$ The quoted authors 
adopted a completely different approach to estimate the viewing angle, based on kinematics. 
$b)$ The difference in age from
short-- to long--period LMC Cepheids, estimated on the basis of period-age relations,
is of the order of 300 Myr, while RSG 
have typically ages of few tens of Myr \citep{bono15}.
Thus, the similarity between the geometrical proprieties of the two tracers
implies that RSGs and Cepheids belong to the same young population.

{\em iv) Comparison with intermediate-age stellar populations}--The LMC viewing angles 
provided by \citet{subramanian13} using RC stars agree quite well with current estimates. 
This is an interesting finding, given that the accuracy of 
individual distances of RC stars is still lively debated in the literature, since 
it might be affected by differences in the underlying stellar populations. 
The LMC viewing angles estimated by \citet{vandermarel01b} using AGB stars show a difference at 
the 2$\sigma$ level. It is not clear whether the difference is mainly caused by 
the LMC area covered by their sample, which extends further than the region were 
Cepheids are located, or by a possible mix of old and intermediate--age  AGB stars.     

{\em v) Comparison with old stellar populations}--Very accurate LMC viewing angles 
have been recently provided by  \citet{deb14}  using a large sample ($\sim$13,000) 
of RR Lyrae stars  
covering a significant fraction of the LMC body. They found a position angle that 
is at least 25 degrees larger than the current one, moreover, the inclination 
angle is at least two degree smaller. This difference taken at face value is 
further confirming that old and young stellar populations in the LMC have 
different radial distributions and likely a different center of mass.    

Similar differences in radial 
distributions have already been found in several nearby dwarf galaxies 
\citep{monelli03,bono10}.
This difference also appears in the chemical composition of the two populations. Indeed, 
\citet{fabrizio15} found evidence that the old and intermediate-age populations 
in the Carina dwarf spheroidal display different mean iron and magnesium abundances.  
Thus suggesting that they experienced different chemical enrichment histories. 
A similar empirical scenario is also disclosed by LMC Cepheids and RR Lyrae stars. 
Recent spectroscopic investigations based on high-resolution spectra indicate that 
the metallicity distribution of LMC Cepheids is centred on [Fe/H]=-0.33 with a 
standard deviation of 0.13 dex \citep{romaniello08}. On the other hand, spectroscopic measurements of 
LMC RR Lyrae based on low-resolution spectra \citep{clementini00}
indicate a  mean [Fe/H]$\sim$-1.5 standard deviation of 0.5 dex.  

The above evidence are further supporting the hypothesis that the LMC old and young 
stellar populations have had significantly different chemical enrichment 
histories.  

In this context it is worth mentioning that the LMC viewing angles provided by 
\citet{vandermarel14} using RGs are quite different when compared with RR Lyrae 
stars. The inclination is more than 1.5$\sigma$ larger, while the position 
angle is significantly smaller. A detailed analysis of the difference is beyond 
the aim of the current investigation.  However, we note that RG are not "pure"   
old tracers, since intermediate-mass stars also contribute to the field population. 

\section{A new reddening map of the LMC disk}

The use of a multi-wavelength fitting of the reddening law to apparent 
distance moduli of extra-galactic Cepheids, to determine their distances 
and reddening, was introduced by \citet{freedman85,freedman91}
and has been recently revised by \citet{rich14}. 

The above method is based on the evidence that the true distance modulus of the 
$i$-th Cepheid belonging to a stellar system can be written in the following form:
\begin{eqnarray}
\label{eq4}
\mu_{0,i}=\mu_{obs,i}(x)+ (a(x) R_V +b(x)) \times  E_i(B-V)
\end{eqnarray}

where $x\equiv\lambda^{-1}$, $a(x)$ and $b(x)$ are the coefficients of the adopted 
reddening law \citep{cardelli89}. 
Fitting this relation to the apparent distance moduli of the same Cepheid estimated 
using different photometric bands and by extrapolating to $x\sim0$, we can determine its 
true distance modulus. Adopting the observed PL relations in the $V,I,J,H$, 
$K_{\rm{S}}$ and $w1$-bands\footnote{$x_V=1.835\mu m^{-1}$; $x_I=1.253\mu m^{-1}$; 
$x_J=0.800\mu m^{-1}$; $x_H=0.606\mu m^{-1}$; $x_K=0.465\mu m^{-1}$;$x_{w1}=0.286\mu m^{-1}$.} bands
we can derive apparent distance moduli at six different wavelengths
for all the Cepheids in our Sample~B.  
The inclusion of the $w1$-band mean magnitudes allow us 
to overcome possible systematics in the extrapolation to extremely long wavelengths.
By assuming $R_V=3.23$  and performing a fit of the above equation we can evaluate 
the true distance modulus and the color excess for individual Cepheids.   

The zero-points of the six observed PL relations were calibrated following the 
same approach used in \citet{inno13}. We adopted nine FU Cepheids for which 
HST parallaxes are available \citep{benedict07,vanleeuwen07}. The FO PL 
relations were calibrated only using Polaris \citep{vanleeuwen07}.

\begin{figure}[!ht]
\includegraphics[width=\columnwidth]{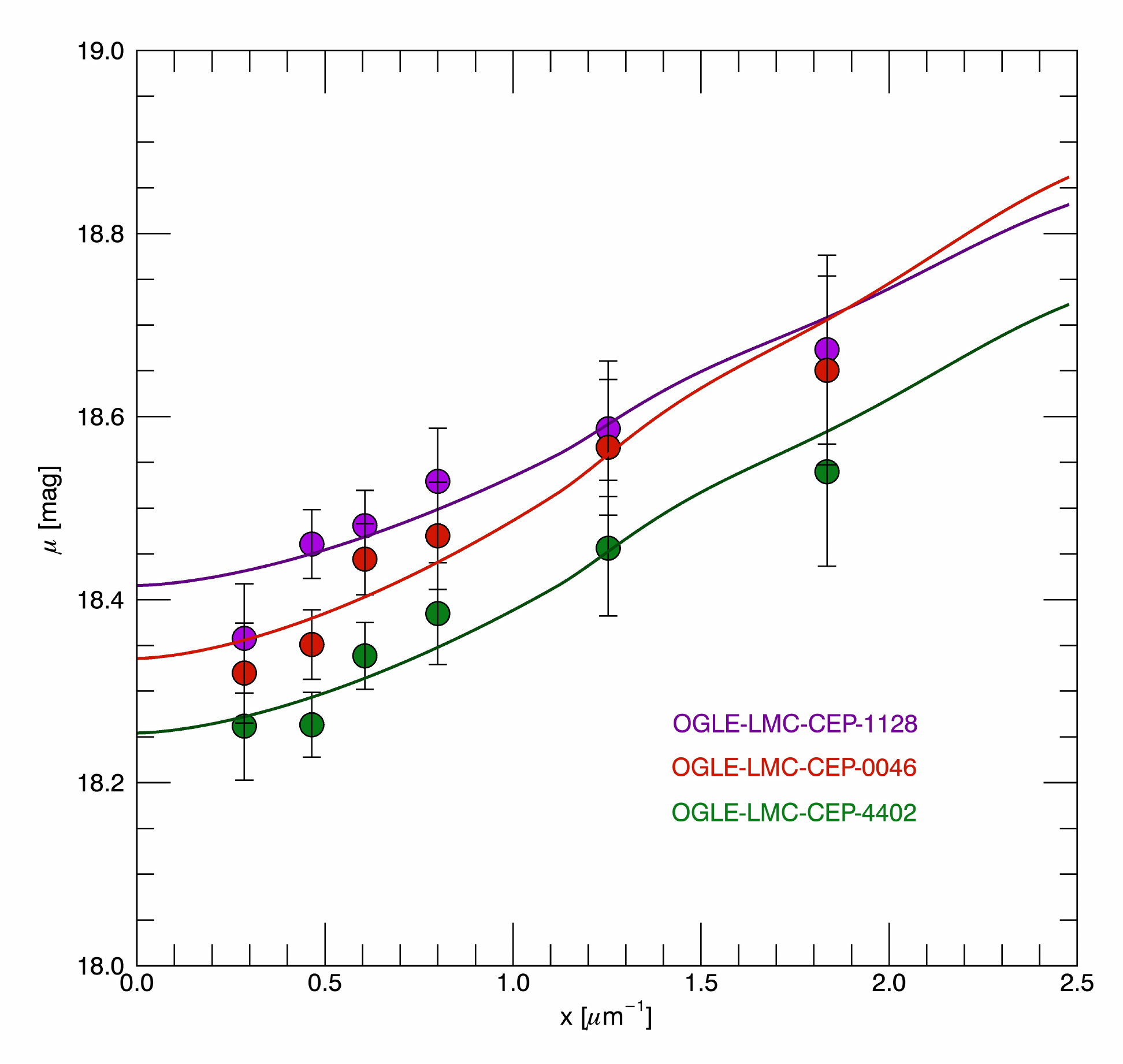}
\caption{Apparent distance moduli (uncorrected for any reddening) for three different Cepheids (CEP-0107, purple; CEP-0683, green; CEP-2337, red) as a function of inverse wavelengths, with the associated error bars. The errors include both measurement and systematic effects.
The best-fit of the reddening law is also showed (solid lines) for each of them (see Section~7 for more details). }
\label{f9}
\end{figure}

The optical and NIR mean magnitude of these calibrating Cepheids have been 
accurately measured \citep{benedict07,fouque07,storm11a}, however, $w1$-band 
light curves for the same Cepheids are not available, since they are saturated 
in the survey. Fortunately enough, the difference in the photometric zero-point 
between the $w1$-band adopted by WISE and the [3.6]-band adopted by SPITZER is 
vanishing (M. Marengo private communication).
Therefore, the $w1$-band FU and FO PL relations were calibrated using the mean [3.6] 
magnitudes for calibrating Cepheids based on SPITZER observations provided 
by \citet{marengo10}.  
Figure~\ref{f9} shows the apparent distance moduli for three selected Cepheids 
in our sample (green, red and magenta dots) as a function of $x$ and the best-fits  
estimated using Eq.~5 (green, red and magenta solid lines). The individual true 
distance moduli and the color excess are labeled together with their errors.
The error bars also account for the systematic
errors related to the position of the Cepheids inside the instability strip 
compared with the ridge line of the adopted PL relation. Individual distances 
based on PL relations rely on the assumption that the width in temperature 
of the instability strip can be neglected \citep{bono02}. On the other hand,
distances based on PLC relations are not affected by this drawback \citep{bono99b,bono02}. 
To quantify this systematic error we adopted the standard deviations of 
theoretical PL relations in the above six bands and they are listed in 
Table~2. This is an upper limit to the standard 
deviations of the PL relations, since predictions  cover the entire 
period range and uniformly fill the instability strip.   

Even a cursory look to the values listed in Table~2 shows that the 
standard deviations, as expected, steadily decrease for increasing 
wavelength. The difference is caused by the fact that cooler Cepheids 
become in NIR and in MIR systematically brighter due to a stronger 
sensitivity of the bolometric correction \citep{bono99,bono99b,bono00}. 
In the case of the $w1$-band we adopted the dispersion around the 
SPITZER [3.6]$\mu$m PL relation provided by \citet{ngeow12}, 
i.e., 0.025 mag.

The approach we adopted to estimate the error budget implies that we 
automatically weight more the apparent distance moduli based on NIR- and 
MIR mean magnitudes than the distances based on optical 
bands. The total error is then propagated on the parameters estimated 
by our least-squares fitting procedure. It is worth noting that current individual 
distance moduli, taking account for systematic errors, have an accuracy better 
than 1\%, while individual extinctions have an accuracy better than 15\%. 

\begin{figure}[!ht]
\begin{center}
\includegraphics[width=\columnwidth]{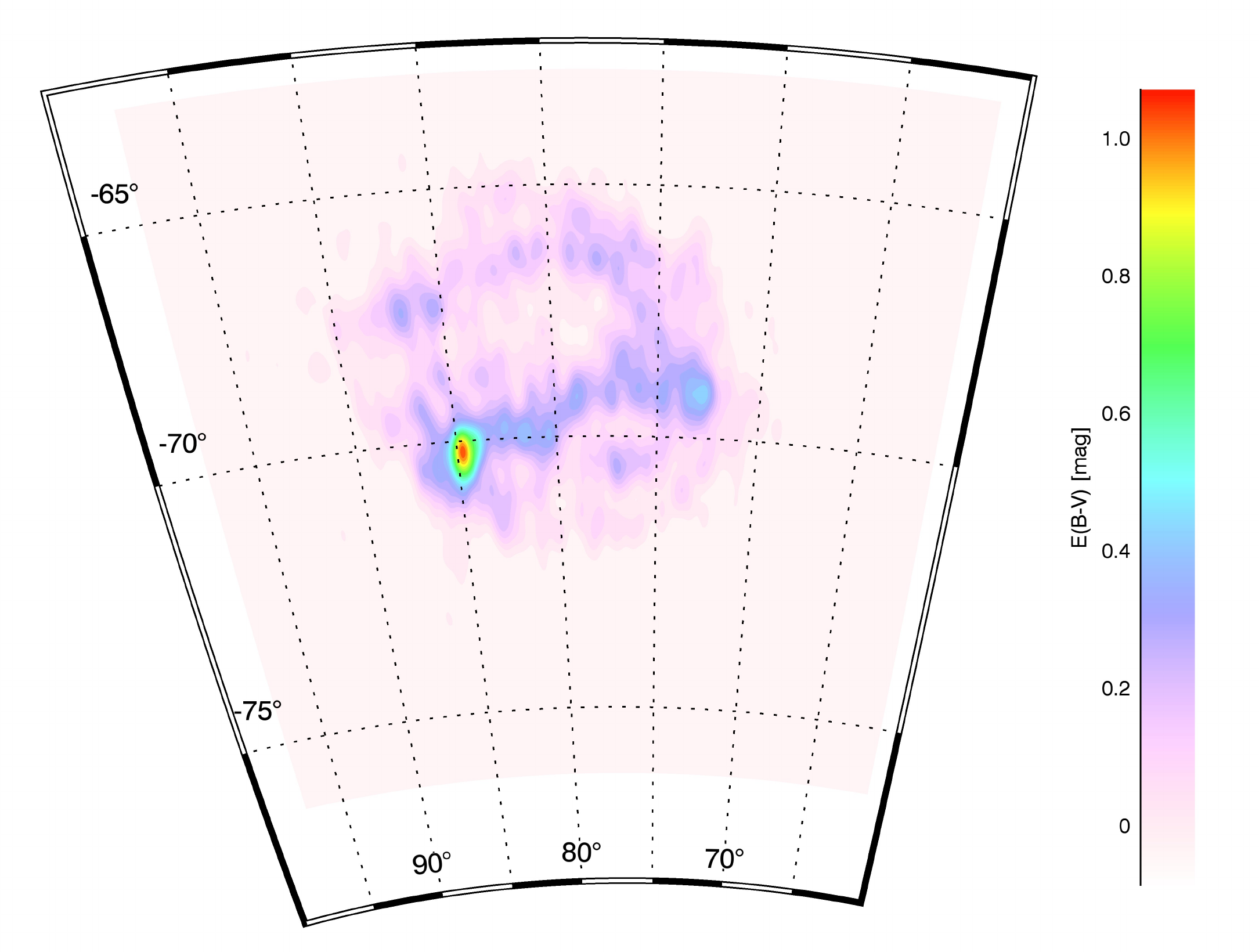}
\caption{Reddening map of the LMC disk derived through the method described in Section~7. 
A Gaussian-smoothing has been applied to obtain the contour map.  
Individual values of the estimated reddening are given in Table~9.
}
\label{f11}
\end{center}
\end{figure}

The individual reddening estimates were smoothed using a Gaussian kernel with 
a $\sigma$ equal to the observed uncertainty. The smoothed reddening distribution 
was adopted to provide the reddening map shown in Figure~\ref{f11}. 
%
The star forming region of 30~Doradus stands out in the reddening map as the most extincted
region on the LMC bar, while on the other side of the bar, the star forming regions 
associated with NGC~1850 and NGC~1858 are also heavily extincted and they can also 
be easily identified. The reddening across the LMC disk is, as expected, quite 
low with the exception of these peculiar regions.

\subsection{Mean distance and reddening to the LMC}

\begin{figure}[!ht]
\includegraphics[width=\columnwidth]{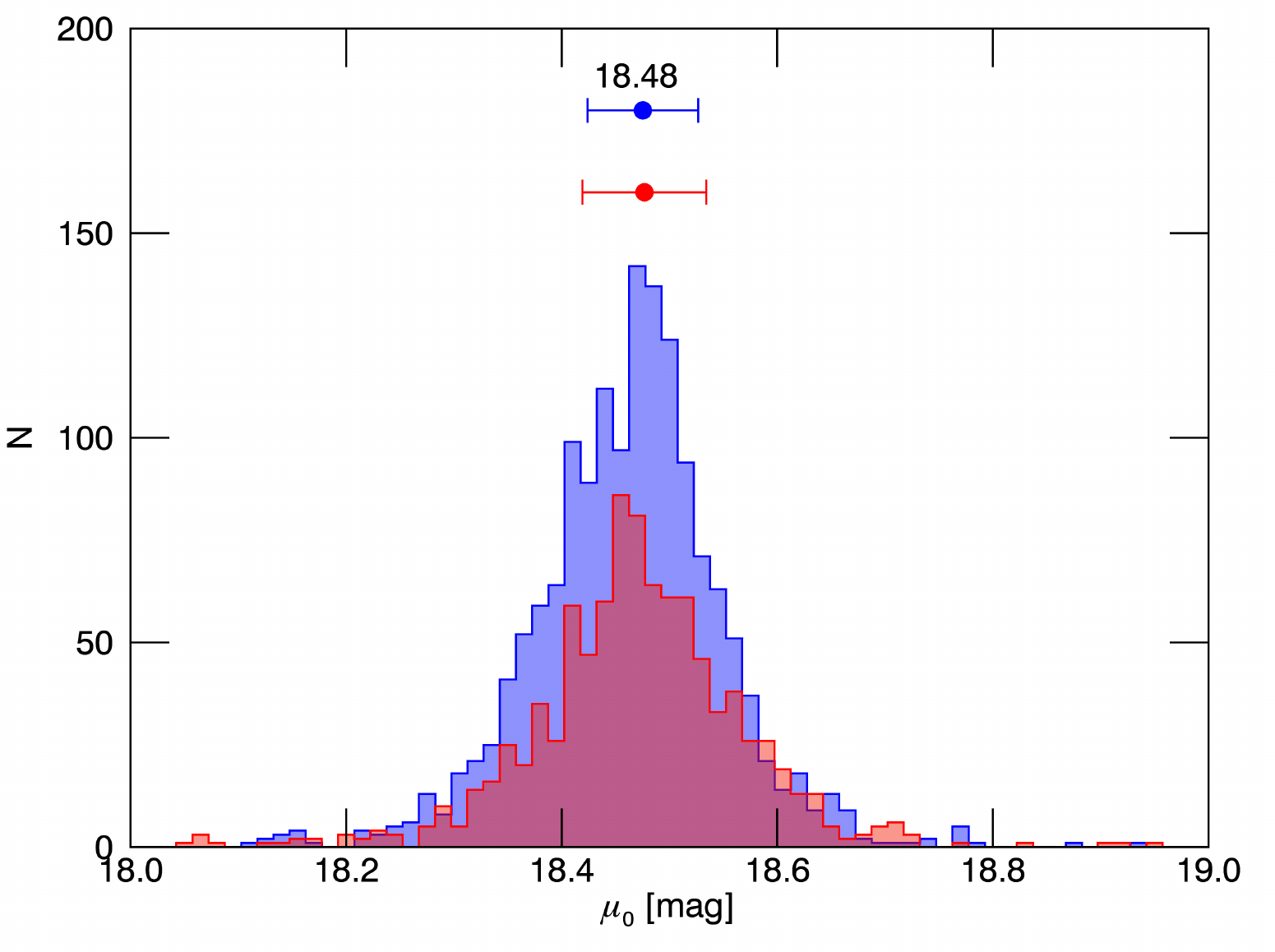}
\caption{Histogram of the true distance moduli $\mu_{0,i}$ distribution for FU (blue bars) and FO (red bars) Cepheids obtained from Equation~\ref{eq4}. 
The blue dot indicates the median distance from the FU Cepheids' distribution, while the error bar indicates the 1$\sigma$ dispersion around the median, $\sigma$=0.10 mag.
The red dots shows the same for FO Cepheids, with $\sigma$=0.11 mag.
We thus find a median distance modulus to the LMC $\mu$=18.48 $\pm$ 0.10 mag for both FU and FO Cepheids.
The above value is in excellent agreement with the mean distance modulus by P13, which is accurate to 2.2\%.
}
\label{f10}
\end{figure}

We also derived the average LMC true distance modulus and reddening.
The statistical error associated to the distance modulus is dominated
by the number of bands available, and in our case it is significantly small, 
and indeed the values range from 0.008 to 0.015 mag. 
Figure~\ref{f10} shows the histogram of the distance moduli distribution
for FU (blue bars) and FO Cepheids (red bars).
We find a median distance modulus
$\mu_0=18.48\pm 0.10$ mag for both FU and FO pulsators,
where the error is given by the standard dispersion around the median.
The current distance moduli based on FU and FO Cepheids are 
in excellent agreement with each others and with the value recently published 
by P13, i.e., 18.493 mag $\pm$ 0.008 (statistical) $\pm$ 0.047 (systematic).
However, note that both distances and
reddening values display neither a symmetric nor a Gaussian distribution.
This means that the median values need to be cautiously treated. 
Thus, we only adopted the median value 
of the entire (FU$+$FO) sample for the purpose of comparing it to 
similar values in the
literature and we found that 
our estimate $E(V-I)=1.265 \times  E(B-V)=0.11 \pm 0.09$ mag, 
is  in excellent agreement (within $1\sigma$) with the value 
estimated by \citet[][hereinafter H11]{haschke11} 
using RC stars ($E(V-I)=0.09 \pm 0.07$ mag),
and RR Lyraes ($E(V-I)=0.11 \pm 0.06$) mag. 

However, individual distances and reddening are 
strongly related to the spatial distribution of the 
Cepheids in the LMC, as discussed in the previous Sections.
Thus, we need to compare individual values and in particular we can compare
$E(V-I)$ values for $\sim$1,000 stars 
located at positions for which reddening from H11 is available.
We download their catalog from the website: \url{http://dc.zah.uni-heidelberg.de/mcx}.
The catalog provides mean reddening values for specific spatial bins whose coordinates
are also available. 
We estimated the mean $E(V-I)$ values in the same spatial bins but from our reddening map.
We first determined the number of Cepheids included in each bin, and 
we only considered bins that contain at least one Cepheid. We found $\sim$500
bins that include from two to eight Cepheids.
Finally, we estimated the difference in reddening from our mean values and the values
by H11 for the same bin.
We found that the difference is smaller than 1$\sigma_{EVI}$ for $\sim$85\% of the bins,
where the error $\sigma_{EVI}$ is given by the sum in quadrature of the error on  
$E(V-I)$ from H11(30--400\%) and from our estimates (0.2--10\%).

Moreover, our new reddening map has two significant advantages when
compared to the map provided by H11: 
$a)$ it covers a double area of the LMC (80 vs 40 square degrees), 
and in particular, it covers the whole disk for the very first time;
$b)$ it is characterised by a much higher accuracy. The total error 
(including systematics) on the reddening is smaller than 10\%, which is one 
order of magnitude better than the typical accuracy of existing reddening maps. 

To further quantify the accuracy of the current reddening estimates, we 
performed a detailed comparison with accurate reddening measurements 
available in the literature. Recently, P13 measured extinction for 
eight double eclipsing binary systems (DEBs) located in the bar and on the left 
arm of the LMC. We performed a beam search of 5' in radius around these systems and computed
the median extinction from the Cepheids included in these radius. 
The comparison between the values based on Cepheids and the ones
obtained by P13 is shown in Table~\ref{tab6}. 
The redding values are in excellent agreement for six out of the eight systems 
for which we found the match.
In two cases (OGLE-LMC-ECL-26122, OGLE-LMC-ECL-09114), our reddening values are 
lower than the ones obtained by P13.  However, on the basis of their individual 
distances these two systems seem to be located behind the plane 
of the LMC disk, and thus they might suffer a higher extinction when compared with 
Cepheids located in the LMC bar. 

\citet{marconi13} find an extinction of $E(V-I)=0.171 \pm 0.015$ mag for the
Cepheid OGLE-LMC-CEP-0227, belonging to a detached double-lined eclipsing binary system
located at ($\alpha$, $\delta$)=(73$^{\circ}$.0654,-70$^{\circ}$.2420).
We found a median extinction $E(V-I)=0.157 \pm 0.001$ mag for the Cepheids
at 5' in radius from this system, thus in perfect agreement with the value 
by \citet{marconi13}.

Recently, \citet{elgueta16} estimates the distance and reddening to the OGLE-LMC-ECL-25658
binary system, which is located at ($\alpha$, $\delta$)=(90.4949,-68.5153).
They found $E(B-V)= 0.091 \pm 0.030$ mag, while we found $E(B-V)= 0.096 \pm 0.005$ mag  
for a Cepheids located at ($\alpha$, $\delta$)=(90.2155,-67.8089)
and $E(B-V)= 0.105 \pm 0.005$ mag for a Cepheids located at 
($\alpha$, $\delta$)=(90.6012,-69.2539).

The comparison with literature values further supports the precision 
and the accuracy of the reddening map we estimated using classical 
Cepheids as tracers of stellar populations in the LMC disk. 

\subsection{Period-Luminosity relations corrected for reddening}
The accurate individual reddening values 
we determined for all the Cepheids in our Sample~B,
allow us to determine new PL relations for the LMC Cepheids
in six different bands. 
We adopted the following absorption coefficients for unit of $E(B-V)$:
A$_I=0.608$,A$_J=0.292$, $A_H=0.181$, $A_{K_S}=0.119$ and $A_{w1}=0.055$,
to transform the $E(B-V)$ values into the absorption in each band.
Thus, we computed the reddening-corrected magnitude for 
each Cepheid in each band and performed
a least-squares fit to determine the PL relations
in the form: $a$+$b \log P$. 
The best-fit parameters and dispersions of the six PL relations for
FU and FO Cepheids are given in Table~\ref{tab7}.
Note that the slopes we find are in agreement within 
1$\sigma$  or better (see e.g. the $J$ band PL)
with the ones from theoretical predictions listed in Table~\ref{tab2_bis}.
Moreover, we also list the PL relations
obtained by adopting the extinction values by
H11 for $\sim$1,120 FU and $\sim$ 780 Cepheids.
The slopes found with the two different values of the reddening
agree inside the error-bar given by the scatter of the relation. 
However, the dispersion around the optical PL relations obtained
by using the H11 correction are a factor 1.5 ($I$) and 2.5 ($V$)
larger than the ones found by using our new correction.
This finding further supports the high accuracy of our new 
reddening map for the LMC disk. 
Moreover, the slopes of the  $J$,$H$ and $K_{\rm{S}}$
PL relations are in excellent agreement with the ones found by P04,
who adopted the reddening computed for 51 Cepheids by \citet{gieren98},
and with the ones by \cite{storm11b}, who used instead a completely independent approach
based on the Baade-Wesselink method. 
\begin{figure}[!ht]
\includegraphics[width=\columnwidth]{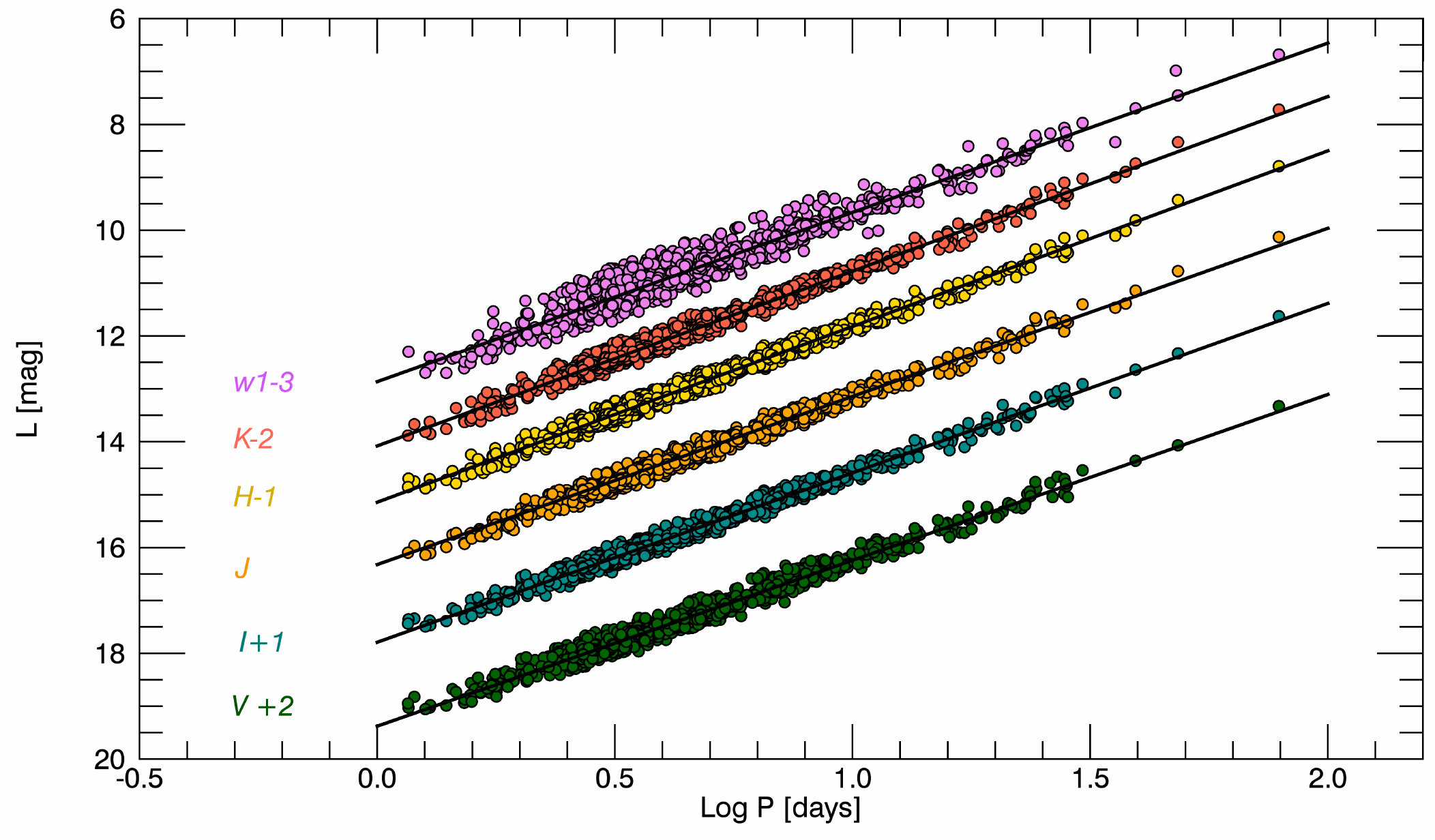}
\caption{PL relations for FU Cepheids in our Sample~B corrected for reddening by using our new reddening map.
The best-fit parameters of the PL relations are given in Table~\ref{tab7}.
The dispersion around each of this relation ranges between 0.09 mag ($V$-band) to 0.17 mag ($w1$-band), 
which is a factor 65\%($V$-band)--10($K_{\rm{S}}$-band)\% smaller than the dispersion when no reddening correction is applied (see Table~\ref{tab7}).  
The PL have been arbitrarily shifted in integer magnitude steps to improve the clarity of the figure (see annotations). 
}
\label{f12}
\end{figure}
%
Finally,the new slopes are also in excellent agreement with the ones found by \citet{macri15}
by adopting the reddening corrections by H11. In fact, the agreement is 
at $0.1\sigma$ level for our relations estimated by using the reddening corrections by H11,
and at $1\sigma$ level when using our new reddening corrections.

\section{Summary and Discussion}

We collected the largest ($\sim$~4,000) sample of optical-NIR-MIR measurements 
for LMC Cepheids. The use of multi-wavelength observations and of accurate 
NIR templates, allowed us to determine 3\%(optical) --15\%(NIR) 
precise individual distances for the entire sample of Cepheids. 
Moreover, we adopted theoretical predictions based on up-to-date pulsation 
models to quantify possible systematic errors on individual Cepheid distances.

We found that individual distances based on optical PW relations
are more affected by systematics (uncertainty on the adopted reddening 
law, intrinsic dispersion), when compared with distances based on 
NIR PW relations.  Using the predicted intrinsic dispersion for the 
PW$_{VI}$ relation ($\sim$0.06 mag), we found that individual distances
 \emph{only} based on optical mean magnitudes cannot have an accuracy better than 7\%. 
On the other hand, the simultaneous use of the three NIR bands: $J$,$H$ and $K_{\rm{S}}$,
allow us to nail down the systematics and to provide individual distances with 
an accuracy better than 1.5 \%. 
However, the uncertainty on the W$_{HJK}$ mean magnitudes due to the photometric errors
($\sigma_{J=17}\sim$0.05--0.15 mag) on single observations effectively limits 
the above accuracy, for some samples, to 15\%.
The error budget on individual Cepheid distances, when moving from optical to 
NIR bands is dominated by different uncertainties (systematics vs measurement 
errors). This gives us the unique opportunity to internally validate distances 
together with their errors and reddening. Our main results are summarised in the following.

$\bullet${\em Viewing angles --} 
We find that the disk of the LMC is oriented with an inclination of
$i=25.05 \pm 0.02$~(stat.)~$\pm 0.55$~(syst.)~deg and  a
position angle of P.A.=150.76$\pm$0.02~(stat.)~$\pm$ 0.07~(syst.).
These values are in excellent agreement with recent estimates based on
stellar tracers of similar age \citep[RSG stars,][]{vandermarel14}.
On the other hand, previous investigations based on Cepheids
found larger inclinations \citep[][P04]{nikolaev04,haschke12} and
smaller position angles \citep[][P04]{haschke12}. The difference is
caused by the different spatial distribution of the adopted Cepheid
samples. The LMC viewing angles depend, due to its non-axisymmetric
shape, on the sky coverage of the adopted stellar tracers. Moreover,
the dependence of the inclination on the distance from the center
of the distribution is mainly due to a limited mapping of the disk.
Using Cepheids that are only located in the central fields, i.e., the
LMC bar according to the definition by \citet{nikolaev04}, we found
$i=35.^{\circ} 8  \pm 5^{\circ}$. Note
that the error reported here and in the following is the difference between the angles found by 
adopting the PW$_{VI}$ and the PW$_{HJK}$ relations. 
If we extend the region covered by Cepheids towards the western part,
i.e., till the edge of the north-western arm, we found $i=30^{\circ}\pm10^{\circ}$, 
which is very close to the values found by \citet{weinberg01} and by \citet{nikolaev04}.

If we exclude the Cepheids located across the bar,  the inclination is 
$\sim$24$^{\circ}$.2, thus perfectly consistent with what we have 
already found using
the entire sample. This finding further supports the evidence that the current
Cepheid sample allow us to precisely determine its geometry, since it traces the
whole LMC disk, i.e., the bar plus the spiral arms.
There is mounting empirical evidence that stellar tracers ranging from 
old, low-mass (RR Lyraes) to intermediate-mass (planetary nebulae, red 
clump, AGB) stars and evolved young massive stars (RSGs) do provide 
different viewing angles.
The difference between old and young stellar tracers indicate that the 
former one is slightly less inclined by $\sim3^{\circ}$ and has a position angle 
$\sim20^{\circ}$ degree larger than those based on Cepheids. 
The above evidence needs to be supported by radial velocity
measurements for large samples of the quoted stellar tracers.

A few years ago, \citet{minniti03} using accurate individual 
distances of 43 LMC RR Lyrae based on the K-band PL relations 
and kinematic measurements, proposed the possible existence 
of a dynamically hot spherical halo surrounding the LMC. 
However, subsequent estimates based on star counts covering 
a broader area up to 20$^{\circ}$ from the LMC center \citep{saha10} 
and on RG kinematics \citep{gallart04,carrera11} did not support this finding.
The possible occurrence of an extended disk is also still controversial 
\citep{majewski09,saha10,besla16}.
The same outcome applies to the possible occurrence
of metallicity gradients among the individual stellar components. The 
possible occurrence of radial gradients in the metallicity distribution 
appears even more promising, since it will allow us to couple the different star 
formation events with their own chemical enrichments and their radial migrations.
The MCs play in this context a crucial role, since the difference in radial 
distance among the different stellar tracers is negligible.

$\bullet${\em Distance to the LMC--}
Taking advantage of the multi-band (optical-NIR-MIR) data-set,
we adopted the reddening-law fitting method \citep{freedman85} to determine 
simultaneously
the true distance modulus and the reddening of the entire Cepheid 
sample.
We take account for both estimate and systematic error on individual 
distance moduli and we found that the final error ranges from 0.1\% to  
0.7\%. We computed the LMC distance distribution and we found that the
median is $\mu_0=18.48\pm 0.10$ mag using both fundamental and first overtone Cepheids.
The above error, estimated as the standard deviation around the median, 
accounts for both statistical and systematic effects, but neglects the error on
the zero-point of the photometric calibration ($\sim$0.02 mag).
The excellent agreement on the distance based on fundamental and first overtone Cepheids
further supports the use of FO Cepheids as solid distance indicators
\citep{bono10,inno13}. Moreover, our estimate of the mean distance to the LMC is also 
in excellent agreement with similar estimates, but based on a smaller Cepheid
sample \citep{inno13}, and with the geometrical distance 
obtained by \citet{pietrzynski13} on the basis of eclipsing binary systems.

$\bullet${\em Reddening towards the LMC disk --}
The reddening-law fitting method provides individual reddening estimates
for each Cepheid in our sample, with an accuracy better than 20\%.
We compared the current reddening values with those available in the
literature and we found that the current reddening map agrees quite well
with the one provided by \citet{haschke11} using RC stars, but it is
one order of magnitude more accurate and a factor of two larger.
We provide the entire Cepheid catalog with mean NIR and MIR magnitudes,
together with the individual distances and extinction values.

We demonstrated that the use of NIR PW relations to determine Cepheids
individual distances is extremely promising. However,
NIR surveys towards the LMC with modest photometric precision 
($\sigma_{J=17}\sim$0.05--0.15 mag, on single observations) 
do not allow us to fully exploit the intrinsic accuracy of NIR
distance diagnostics.
However, accurate NIR templates allow us to use highly-accurate
single-epoch photometric measurements available in the literature
to match the precision typical of distance determinations based
on optical bands.

The current approach based on measurements ranging from optical to mid-infrared
observations of classical Cepheids, appears very promising for
accurate individual distance determinations and paves the way to
accurate estimates of their intrinsic properties as a function of
the radial distribution.

We will complement the accurate information on the three-dimensional distribution
presented here with individual radial velocities and chemical abundances
for a significant fraction of the Cepheids in our sample. The kinematic and 
the chemical tagging of a significant fraction of LMC Cepheids will allow us 
to further constrain the physical proprieties of the young stellar population 
in the LMC disk.

\acknowledgments
This work was supported by Sonderforschungsbereich SFB 881
"The Milky Way System" (subproject C9) of the German Research Foundation (DFG).
The OGLE project has received funding from the National Science Centre,
Poland, grant MAESTRO 2014/14/A/ST9/00121 to AU.
One of us (G.B.) thanks the Japan Society for the Promotion of Science
for a research grant (L15518).
It is also a pleasure to thank the anonymous referee for his/her
supportive attitude and insightful suggestions that helped us to improve
the readability of the paper.


\clearpage

\clearpage

\begin{center}
\begin{deluxetable}{llccccl}
\tabletypesize{\scriptsize}
\tablewidth{0pt}

\tablecaption{Sub-Samples adopted \label{tab1}}

\tablehead{
\colhead{SAMPLE}&
\colhead{FU}&
\colhead{FU Period range}&
\colhead{FO}&
\colhead{FO Period range}&
\colhead{Use of the Templates}&
\colhead{$\sigma_{Ph}$@ $J$=16 mag}
\\
\colhead{}&
\colhead{\#}&
\colhead{[days]}&
\colhead{\#}&
\colhead{[days]}&
\colhead{}&
\colhead{[mag]}\\
}

\startdata
VMC &  142 & 1--20.1  & 118 & 0.7--5 &Y & 0.03\\
IRSF & 1418  & 1--28.8  & 846 & 0.7--6 &Y & 0.03\\
CPAPIER & 60  &1--28.2 & 48 & 0.7--5 & Y & 0.10\\
2MASS & 560  & 1--28.0 & 525 & 0.7--5 &Y & 0.10\\
P04 & 65  & 2--99.2 &  -- &  -- & N & 0.02\\
WISE & 1,557  & 2--63 &  1,086 &   0.7--5 & N & 0.02\\
&&&&&&\\
TOTAL Sample~A\tablenotemark{a} & 2,245 & 1--99.2 & 1,537 & 0.7 --6 & & \\
TOTAL  Sample~B\tablenotemark{b}& 1,557 & 1--99.2 & 1,086 & 0.7 --6 & & \\
\enddata    
\tablenotetext{a}{optical-NIR dataset}
\tablenotetext{b}{optical-NIR-MIR dataset}
\end{deluxetable} 
\end{center}

%

\begin{center}

\begin{deluxetable}{lccc}
\tabletypesize{\scriptsize}
\tablewidth{0pt}

\tablecaption{Theoretical NIR and Optical-NIR PW relations for LMC Cepheids in the form: $a + b \log P$ \label{tab2}}

	
\tablehead{
\colhead{Wesenheit definition}&
\colhead{a  $\pm \sigma_a$}&
\colhead{b   $\pm \sigma_b$}&
\colhead{$\sigma$\tablenotetext{a}}
}
\startdata
W$_{JH}$ =   $H$ - 1.63  $\times$ ( $J- H$)                    & -3.066 $\pm$ 0.002 & -3.464 $\pm$ 0.002  & 0.011\tablenotemark{b}\\
W$_{HJK}$ =  $H$ - 1.046 $\times$ ( $J- K_{\rm{S}}$)     & -2.930 $\pm$ 0.001 & -3.394 $\pm$ 0.001  &   0.027\\
W$_{JK}$ =   $K_{\rm{S}}$ - 0.69  $\times$ ( $J- K_{\rm{S}}$)          & -2.849 $\pm$ 0.001 & -3.351 $\pm$ 0.001  & 0.038\\
W$_{IH}$ =   $H$ - 0.42  $\times$ ( $I- H$)                   & -2.869 $\pm$ 0.001 & -3.348 $\pm$ 0.001  & 0.038\\
W$_{HIK}$ =   $H$ - 0.370 $\times$ ( $I- K_{\rm{S}}$)    & -2.849 $\pm$ 0.001 & -3.338 $\pm$ 0.001  &   0.041\\
W$_{VH}$ =    $H$ - 0.22  $\times$ ( $V- H$)           & -2.866 $\pm$ 0.001 & -3.339 $\pm$ 0.001  & 0.041\\
W$_{HVI}$ =   $H$ - 0.461 $\times$ ( $V- I$)  & -2.862 $\pm$ 0.001 & -3.330 $\pm$ 0.001 &  0.044\\
W$_{KIH}$ =   $K_{\rm{S}}$ - 0.279 $\times$ ( $I- H$)     & -2.809 $\pm$ 0.001 & -3.321 $\pm$ 0.001  &   0.045\\
W$_{KVH}$ =   $K_{\rm{S}}$ - 0.145 $\times$ ( $V- H$)   & -2.805 $\pm$ 0.001 & -3.315 $\pm$ 0.001  &   0.047\\
W$_{IK}$ =   $K_{\rm{S}}$ - 0.24  $\times$ ( $I- K_{\rm{S}}$)   & -2.791 $\pm$ 0.001 & -3.314 $\pm$ 0.001  & 0.047\\
W$_{VK}$ =    $K_{\rm{S}}$ - 0.13  $\times$ ( $V- K_{\rm{S}}$) & -2.792 $\pm$ 0.001 & -3.309 $\pm$ 0.001  & 0.049\\
W$_{KVI}$ =   $K_{\rm{S}}$ - 0.304 $\times$ ( $V- I$)   & -2.778 $\pm$ 0.001 & -3.285 $\pm$ 0.001  &   0.053\\
W$_{JIH}$ =   $J$ - 0.684 $\times$ ( $I- H$)                     & -2.753 $\pm$ 0.001 & -3.278 $\pm$ 0.001  &   0.055\\
W$_{VI}$ =   $I$ - 1.55  $\times$ ( $V- I$)               & -2.838 $\pm$ 0.002 & -3.286 $\pm$ 0.002  & 0.058\\
W$_{HK}$ =    $K_{\rm{S}}$ - 1.92  $\times$ ( $H- K_{\rm{S}}$)  & -2.689 $\pm$ 0.002 & -3.268 $\pm$ 0.002  & 0.059\\
W$_{JVK}$ =   $J$ - 0.331 $\times$ ( $V- K_{\rm{S}}$)      & -2.724 $\pm$ 0.001 & -3.255 $\pm$ 0.001  &   0.062\\
W$_{JVI}$ =   $J$ - 0.745 $\times$ ( $V- I$)    & -2.737 $\pm$ 0.001 & -3.248 $\pm$ 0.001  &   0.065\\
W$_{VJ}$ =   $J$ - 0.41  $\times$ ( $V- J$)               & -2.693 $\pm$ 0.001 & -3.231 $\pm$ 0.001  & 0.068\\
W$_{IJ}$ =   $J$ - 0.92  $\times$ ( $I- J$    )                   & -2.642 $\pm$ 0.001 & -3.212 $\pm$ 0.001  & 0.072\\
W$_{HVK} $=   $H$ - 1.135 $\times$ ( $V- K_{\rm{S}}$)  & -3.939 $\pm$ 0.001 & -3.942 $\pm$ 0.001  &   0.092\\
\enddata

\tablenotetext{a}{The PW relations are listed in order of ascending dispersion}
\tablenotetext{b}{The PW$_{JH}$ relation shows the smallest intrinsic dispersion. However,
we did not adopt this relation because of the coefficient significantly larger than 1 ($\frac{A_H}{E(J-H)}$=1.63) in the Wesenheit definition. In fact, the photometric error on the mean color $E(J-H)$ is also multiplied by the same factor, 
resulting on larger errors on the final Wesenheit magnitude.}

\end{deluxetable} 
\end{center}

%
\clearpage
\begin{center}

\begin{deluxetable}{lcccc}
\tabletypesize{\scriptsize}
\tablewidth{0pt}

\tablecaption{Observed NIR and Optical-NIR PW relations for LMC Cepheids in the form: $a + b \log P$  \label{tab3}}

	
\tablehead{
\colhead{Wesenheit definition}&
\colhead{a  $\pm \sigma_a$}&
\colhead{b   $\pm \sigma_b$}&
\colhead{Num}&
\colhead{$\sigma$}
}
\startdata
\multicolumn{5}{c}{FU CEPHEIDS} \\
\\
W$_{JH}$ =   $H$ - 1.63  $\times$ ( $J- H$)                    & 15.677 $\pm$ 0.003 & -3.377 $\pm$ 0.005 &         2159  & 0.12\\
W$_{HJK}$ =  $H$ - 1.046 $\times$ ( $J- K_{\rm{S}}$)     & 15.788 $\pm$ 0.003 & -3.357 $\pm$ 0.004 &         2164 &  0.10\\
W$_{JK}$ =   $K_{\rm{S}}$ - 0.69  $\times$ ( $J- K_{\rm{S}}$)          & 15.846 $\pm$ 0.002 & -3.331 $\pm$ 0.003 &         2149  & 0.11\\
W$_{IH}$ =   $H$ - 0.42  $\times$ ( $I- H$)                   & 15.831 $\pm$ 0.002 & -3.334 $\pm$ 0.003 &         2168  & 0.09\\
W$_{HIK}$ =   $H$ - 0.370 $\times$ ( $I- K_{\rm{S}}$)    & 15.849 $\pm$ 0.002 & -3.335 $\pm$ 0.003 &         2173 &  0.09\\
W$_{VH}$ =    $H$ - 0.22  $\times$ ( $V- H$)           & 15.837 $\pm$ 0.002 & -3.331 $\pm$ 0.002 &         2168  & 0.09\\
W$_{HVI}$ =   $H$ - 0.461 $\times$ ( $V- I$)  & 15.847 $\pm$ 0.002 & -3.330 $\pm$ 0.002 &         2170 &  0.08\\
W$_{KIH}$ =   $K_{\rm{S}}$ - 0.279 $\times$ ( $I- H$)     & 15.876 $\pm$ 0.002 & -3.321 $\pm$ 0.003 &         2161 &  0.09\\
W$_{KVH}$ =   $K_{\rm{S}}$ - 0.145 $\times$ ( $V- H$)   & 15.881 $\pm$ 0.002 & -3.318 $\pm$ 0.003 &         2166 &  0.09\\
W$_{IK}$ =   $K_{\rm{S}}$ - 0.24  $\times$ ( $I- K_{\rm{S}}$)   & 15.887 $\pm$ 0.002 & -3.312 $\pm$ 0.003 &         2164  & 0.09\\
W$_{VK}$ =    $K_{\rm{S}}$ - 0.13  $\times$ ( $V- K_{\rm{S}}$) & 15.894 $\pm$ 0.002 & -3.314 $\pm$ 0.002 &         2170  & 0.09\\
W$_{KVI}$ =   $K_{\rm{S}}$ - 0.304 $\times$ ( $V- I$)   & 15.944 $\pm$ 0.002 & -3.291 $\pm$ 0.002 &         2173 &  0.09\\
W$_{JIH}$ =   $J$ - 0.684 $\times$ ( $I- H$)                     & 15.917 $\pm$ 0.002 & -3.305 $\pm$ 0.003 &         2175 &  0.10\\
W$_{VI}$ =   $I$ - 1.55  $\times$ ( $V- I$)               & 15.897 $\pm$ 0.001 & -3.327 $\pm$ 0.001 &         2168  & 0.08\\
W$_{HK}$ =    $K_{\rm{S}}$ - 1.92  $\times$ ( $H- K_{\rm{S}}$)  & 15.979 $\pm$ 0.004 & -3.308 $\pm$ 0.005 &         2114  & 0.15\\
W$_{JVK}$ =   $J$ - 0.331 $\times$ ( $V- K_{\rm{S}}$)      & 15.947 $\pm$ 0.002 & -3.294 $\pm$ 0.003 &         2169 &  0.09\\
W$_{JVI}$ =   $J$ - 0.745 $\times$ ( $V- I$)    & 15.948 $\pm$ 0.001 & -3.299 $\pm$ 0.001 &         2175 &  0.08\\
W$_{VJ}$ =   $J$ - 0.41  $\times$ ( $V- J$)               & 15.971 $\pm$ 0.002 & -3.284 $\pm$ 0.002 &         2172  & 0.10\\
W$_{IJ}$ =   $J$ - 0.92  $\times$ ( $I- J$    )                   & 15.997 $\pm$ 0.002 & -3.269 $\pm$ 0.003 &         2168  & 0.11\\
W$_{HVK}$ =   $H$ - 1.135 $\times$ ( $V- K_{\rm{S}}$)  & 14.537 $\pm$ 0.003 & -3.813 $\pm$ 0.004 &         2147 &  0.19\\
W$_{Jw1}$ =   $w1$ - 0.23  $\times$ ( $J- w1$    )                   & 15.756 $\pm$ 0.007 & -3.199 $\pm$ 0.008 &         1489  & 0.20\\
W$_{Hw1}$ =   $w1$ - 0.43  $\times$ ( $H- w1$    )                   & 15.767 $\pm$ 0.007 & -3.167 $\pm$ 0.008 &         1489  & 0.23\\
W$_{Kw1}$ =   $w1$ - 0.86  $\times$ ( $K_{\rm{S}}- w1$    )                   & 15.708 $\pm$ 0.007 & -3.123 $\pm$ 0.008 &         1483  & 0.29\\
W$_{Iw1}$ =   $w1$ - 0.10  $\times$ ( $I- w1$    )                   & 15.791 $\pm$ 0.007 & -3.217 $\pm$ 0.008 &         1486  & 0.18\\
W$_{Vw1}$ =   $w1$ - 0.06  $\times$ ( $V- w1$    )                   & 15.792 $\pm$ 0.007 & -3.218 $\pm$ 0.008 &         1488  & 0.18\\
    \hline
		    \\
\multicolumn{5}{c}{FO CEPHEIDS} \\
\\
W$_{JH}$ =   $H$ - 1.63  $\times$ ( $J- H$)                    & 15.176 $\pm$ 0.004 & -3.458 $\pm$ 0.011 &         1505  & 0.15\\
W$_{HJK}$ =  $H$ - 1.046 $\times$ ( $J- K_{\rm{S}}$)     & 15.258 $\pm$ 0.004 & -3.382 $\pm$ 0.010 &         1539 &  0.14\\
W$_{JK}$ =   $K_{\rm{S}}$ - 0.69  $\times$ ( $J- K_{\rm{S}}$)          & 15.305 $\pm$ 0.003 & -3.323 $\pm$ 0.007 &         1526  & 0.16\\
W$_{IH}$ =   $H$ - 0.42  $\times$ ( $I- H$)                   & 15.316 $\pm$ 0.003 & -3.430 $\pm$ 0.007 &         1543  & 0.13\\
W$_{HIK}$ =   $H$ - 0.370 $\times$ ( $I- K_{\rm{S}}$)    & 15.328 $\pm$ 0.003 & -3.418 $\pm$ 0.008 &         1534 &  0.12\\
W$_{VH}$ =    $H$ - 0.22  $\times$ ( $V- H$)           & 15.328 $\pm$ 0.002 & -3.438 $\pm$ 0.006 &         1544  & 0.12\\
W$_{HVI}$ =   $H$ - 0.461 $\times$ ( $V- I$)  & 15.340 $\pm$ 0.002 & -3.440 $\pm$ 0.006 &         1544 &  0.10\\
W$_{KIH}$ =   $K_{\rm{S}}$ - 0.279 $\times$ ( $I- H$)     & 15.343 $\pm$ 0.003 & -3.363 $\pm$ 0.007 &         1524 &  0.13\\
W$_{KVH}$ =   $K_{\rm{S}}$ - 0.145 $\times$ ( $V- H$)   & 15.350 $\pm$ 0.003 & -3.362 $\pm$ 0.007 &         1524 &  0.13\\
W$_{IK}$ =   $K_{\rm{S}}$ - 0.24  $\times$ ( $I- K_{\rm{S}}$)   & 15.358 $\pm$ 0.002 & -3.357 $\pm$ 0.006 &         1522  & 0.14\\
W$_{VK}$ =    $K_{\rm{S}}$ - 0.13  $\times$ ( $V- K_{\rm{S}}$) & 15.364 $\pm$ 0.002 & -3.358 $\pm$ 0.006 &         1521  & 0.13\\
W$_{KVI}$ =   $K_{\rm{S}}$ - 0.304 $\times$ ( $V- I$)   & 15.440 $\pm$ 0.002 & -3.421 $\pm$ 0.006 &         1549 &  0.11\\
W$_{JIH}$ =   $J$ - 0.684 $\times$ ( $I- H$)                     & 15.402 $\pm$ 0.002 & -3.427 $\pm$ 0.006 &         1562 &  0.14\\
W$_{VI}$ =   $I$ - 1.55  $\times$ ( $V- I$)               & 15.394 $\pm$ 0.001 & -3.434 $\pm$ 0.001 &         1554  & 0.08\\
W$_{HK}$ =    $K_{\rm{S}}$ - 1.92  $\times$ ( $H- K_{\rm{S}}$)  & 15.421 $\pm$ 0.006 & -3.281 $\pm$ 0.015 &         1513  & 0.24\\
W$_{JVK}$ =   $J$ - 0.331 $\times$ ( $V- K_{\rm{S}}$)      & 15.426 $\pm$ 0.002 & -3.400 $\pm$ 0.005 &         1558 &  0.12\\
W$_{JVI}$ =   $J$ - 0.745 $\times$ ( $V- I$)    & 15.438 $\pm$ 0.001 & -3.427 $\pm$ 0.004 &         1566 &  0.11\\
W$_{VJ}$ =   $J$ - 0.41  $\times$ ( $V- J$)               & 15.458 $\pm$ 0.002 & -3.416 $\pm$ 0.004 &         1565  & 0.13\\
W$_{IJ}$ =   $J$ - 0.92  $\times$ ( $I- J$    )                   & 15.478 $\pm$ 0.002 & -3.402 $\pm$ 0.005 &         1568  & 0.16\\
W$_{HVK}$ =   $H$ - 1.135 $\times$ ( $V- K_{\rm{S}}$)  & 13.948 $\pm$ 0.004 & -3.503 $\pm$ 0.010 &         1543 &  0.25\\
W$_{Jw1}$ =   $w1$ - 0.23  $\times$ ( $J- w1$)                   & 15.253 $\pm$ 0.003 & -3.331 $\pm$ 0.007 &         1036  & 0.22\\
W$_{Hw1}$ =   $w1$ - 0.43  $\times$ ( $H- w1$)                   & 15.293 $\pm$ 0.006 & -3.240 $\pm$ 0.014 &         1039  & 0.26\\
W$_{Kw1}$ =   $w1$ - 0.86  $\times$ ( $K_{\rm{S}}- w1$)                   & 15.212 $\pm$ 0.004 & -3.320 $\pm$ 0.010 &         1038  & 0.34\\
W$_{Iw1}$ =   $w1$ - 0.10  $\times$ ( $I- w1$)                   & 15.250 $\pm$ 0.001 & -3.319 $\pm$ 0.003 &         1038  & 0.20\\
W$_{Vw1}$ =   $w1$ - 0.06  $\times$ ( $V- w1$)                   & 15.252 $\pm$ 0.001 & -3.318 $\pm$ 0.003 &         1037  & 0.19\\
\enddata

\end{deluxetable} 
\end{center}

%
\clearpage
\begin{center}
\begin{deluxetable}{lccrr}
\tabletypesize{\scriptsize}
\tablewidth{0pt}

\tablecaption{List of the values adopted as the center of the distribution for LMC Cepheids \label{tab4}}

\tablehead{
\colhead{Definition}&
\colhead{ $\alpha_0$ }&
\colhead{$\delta_0$ }&
\colhead{Reference}&
\colhead{ID}\\
\colhead{}&
\colhead{[ddeg] }&
\colhead{[ddeg] }&
\colhead{}&
\colhead{}\\
}
\startdata

Cepheids centroid & 80.78 & -69.30 & This work & CI \\
HI rotation center & 79.40 &-69.03 &\citet{HI98}& CII \\
optical center &79.91& -69.45 & \citet{devauc72}&  CIII\\
Cepheids geometrical center &80.40&-69.00&  \citet{nikolaev04}& CIV \\
NIR isophote center &81.28&-69.78 &  \citet{vandermarel01b}& CV \\
\enddata    
\end{deluxetable} 
\end{center}

 \clearpage  
 \begin{turnpage}
\begin{deluxetable}{lccllc}
\tabletypesize{\scriptsize}
\tablecaption{Inclination and Position Angle for the LMC \label{tab5}}
\tablehead{
\colhead{Tracers}&
\colhead{P.A.}&
\colhead{$i$}&
\colhead{Reference}&
\colhead{Adopted Center}&
\colhead{sky coverage}\\
\colhead{}&
\colhead{[ddeg]}&
\colhead{[ddeg]}&
\colhead{}&
\colhead{(ID)}&
\colhead{[deg $\times$ deg]}\\
}

\startdata

Cepheids, PW$_{VI}$ & 150.72 $\pm$ 0.02 & 24.46 $\pm$  0.01  & This work & CI & 11.2$\times$14 \\
FU Cepheids, PW$_{VI}$ & 150.79 $\pm$ 0.03 & 24.22 $\pm$ 0.01& This work & CI  &  11$\times$11\\
FO Cepheids, PW$_{VI}$ &  150.68 $\pm$ 0.03 &  24.58 $\pm$ 0.02 & This work & CI   &  11.2$\times$14\\
Cepheids, PW$_{VI}$ & 151.89 $\pm$  0.02  & 24.57 $\pm$  0.01 & This work &  CII  & 11.2$\times$14\\
Cepheids, PW$_{VI}$ & 151.53 $\pm$  0.02  & 24.60 $\pm$  0.01 & This work &  CIII & 11.2$\times$14 \\
Cepheids, PW$_{VI}$ & 151.32 $\pm$  0.02  & 24.56 $\pm$  0.01 & This work &  CIV & 11.2$\times$14 \\
Cepheids, PW$_{VI}$ & 150.38 $\pm$  0.02  & 24.62 $\pm$  0.01 & This work &  CV  & 11.2$\times$14\\

Cepheids, PW$_{HJK}$ & 150.79 $\pm$  0.02  & 25.56 $\pm$  0.01 & This work & CI & 11.2$\times$14 \\
 FU Cepheids, PW$_{HJK}$ & 150.77 $\pm$ 0.03 & 23.42 $\pm$ 0.02 & This work & CI  & 11$\times$11 \\
 FO Cepheids, PW$_{HJK}$ & 150.67 $\pm$ 0.03 & 27.54 $\pm$ 0.02 & This work & CI  & 11.2$\times$14\\

Cepheids, PW$_{HJK}$ & 150.60 $\pm$  0.02 & 25.16 $\pm$   0.02 & This work &  CII  & 11.2$\times$14\\
Cepheids, PW$_{HJK}$ & 151.61 $\pm$  0.02 & 25.07 $\pm$   0.02 & This work &  CIII  & 11.2$\times$14\\
Cepheids, PW$_{HJK}$ & 150.26 $\pm$  0.02 & 25.46 $\pm$   0.02 & This work &  CIV  & 11.2$\times$14\\
Cepheids, PW$_{HJK}$ & 151.95 $\pm$  0.02 & 25.41 $\pm$   0.02 & This work &  CV & 11.2$\times$14\\
Cepheids, best value & 150.76 $\pm$  0.02 $\pm$ 0.07 & 25.05 $\pm$   0.02 $\pm$ 0.55 & This work & CI  & 11.2$\times$14\\
& &  & & & \\
Cepheids, PW$_{VI}$ & 151.4 $\pm$ 1.5 & 24.2 $\pm$ 0.6 & \cite{jac16} &  CII  &  11.2$\times$14\\
Cepheids, PL($V$,$I$),  & 116 $\pm$ 18 & 32 $\pm$ 4 & \citet{haschke12} & CV &  8$\times$ 6.5\\
Cepheids, PL($V$,$R$,$J$,$H$,$K$) & 150.2 $\pm$ 2.4 & 31$\pm$ 1 & \citet{nikolaev04} & CIV & 8$\times$7\\
Cepheids, PL($J$,$H$,$K$) & 127 $\pm$ 10 & 27 $\pm$ 6 & P04 & $\cdots$ &  6.5$\times$ 6.5\\ 
& &  & & & \\
RSG\tablenotemark{a}, 3D-kinematics & 154.5 $\pm$ 2.1 & 26.2 $\pm$ 5.9 & \citet{vandermarel14} & CII & 6.5$\times$8\\  
AGB\tablenotemark{b}, kinematics & 122.5 $\pm$ 8.3 & 34.7 $\pm$ 6.2 & \citet{vandermarel01b} & CV & $\cdots$\\ 
AGB\tablenotemark{b}, kinematics & 142 $\pm$ 5 &$\cdots$ & \citet{olsen11} & (81.9, -69.87) &  8$\times$4\\ 

RC & 148.3 $\pm$ 3.8 & 26.6 $\pm$ 1.3  & \citet{subramanian13}& CIII &  8$\times$6.5\\
RGs & 122$\pm$ 8 & $\cdots$  & \citet{cioni00} & CV &  19.9$\times$16\\ 
RGs\tablenotemark{c}, 3D-kinematics & 139.1 $\pm$ 4.1 & 34.0 $\pm$ 7.0 & \citet{vandermarel14}& CII& 6.5$\times$ 8\\ 
RRab Lyrae & 175.22 $\pm$ 0.01 & 22.25 $\pm$ 0.01 & \citet{deb14} & CIII &  8$\times$5 \\
HI & 168 $\pm$ 1 &   22 $\pm$ 6 & \citet{HI98} &CII &  8$\times$4\\ 
HI & 126 $\pm$ 23 &    $\cdots$ & \citet{indu15} &CII &  20$\times$20\\ 
Isophotes & 170$\pm$ 5 &   27 $\pm$ 2 & \citet{devauc72}&CIII & 16$\times$17\\

\enddata

\tablenotetext{a}{Red supergiant stars, typical ages $\sim$10--50 Myr}
\tablenotetext{b}{Asymptotic giant branch stars, typical ages of $\sim$100 Myr -- 10 Gyr}
\tablenotetext{c}{Red giants, typical ages of $\sim$ 1--12 Gyr}

\end{deluxetable} 
\end{turnpage}
\clearpage

%

\begin{center}

\begin{deluxetable}{lccc}
\tabletypesize{\scriptsize}
\tablewidth{0pt}

\tablecaption{PL theoretical relations for LMC Cepheids in the form: $a + b \log P$ \label{tab2_bis}}

	
\tablehead{
\colhead{Band}&
\colhead{a  $\pm \sigma_a$}&
\colhead{b   $\pm \sigma_b$}&
\colhead{$\sigma$}
}
\startdata

$V$  & -1.447 $\pm$ 0.0004 & -2.605 $\pm$ 0.0004 &       0.203\\
$I$  & -1.987 $\pm$ 0.0004 & -2.879 $\pm$ 0.0004 &         0.145\\
$J$  & -2.324 $\pm$ 0.0004 & -3.057 $\pm$ 0.0004 &         0.105\\
$H$  & -2.610 $\pm$ 0.0004 & -3.207 $\pm$ 0.0004 &          0.070\\
$K_{\rm{S}}$  & -2.636 $\pm$ 0.0004 & -3.229 $\pm$ 0.0004 &      0.066\\
\enddata

\end{deluxetable} 
\end{center}

%
\clearpage
\begin{center}
\begin{deluxetable}{llccc}
\tabletypesize{\scriptsize}
\tablewidth{0pt}

\tablecaption{Comparison of reddening values from Detached Eclipsing Binary Systems (DEBs) by \citet{pietrzynski13} and values from our new reddening map.\label{tab6}}

\tablehead{
\colhead{DES Name}&
\colhead{Ra}&
\colhead{Dec}&
\colhead{$E(B-V)_{DEBs}$}&
\colhead{$E(B-V)_{CEP}$}
\\
\colhead{}&
\colhead{[ddeg]}&
\colhead{[ddeg]}&
\colhead{[mag]}&
\colhead{[mag]}
\\
}

\startdata

OGLE-LMC-ECL-10567 & 78.50788   & -68.6884 & 0.10 $\pm$ 0.02 & 0.11 $\pm$ 0.02 \\
OGLE-LMC-ECL-26122  & 78.52520    & -69.2658 & 0.14 $\pm$ 0.02 & 0.10 $\pm$ 0.01 \\
OGLE-LMC-ECL-09114  & 77.58180    & -68.9701 & 0.16 $\pm$ 0.02 & 0.10 $\pm$ 0.02 \\
OGLE-LMC-ECL-06575  & 76.13696  & -69.3475 & 0.11 $\pm$ 0.02 & 0.11 $\pm$ 0.02 \\
OGLE-LMC-ECL-01866  & 73.06367  & -68.3195 & 0.12 $\pm$ 0.02 & 0.10 $\pm$ 0.01\\
OGLE-LMC-ECL-03160  & 73.96450   & -68.6633 & 0.12 $\pm$ 0.02 & 0.12 $\pm$ 0.02\\
OGLE-LMC-ECL-15260  & 81.35692   & -69.5513 & 0.10 $\pm$ 0.02 & 0.08 $\pm$ 0.03\\

\enddata    
\end{deluxetable} 
\end{center}


\begin{deluxetable}{lcccc}
\tabletypesize{\scriptsize}

\tablecaption{Optical, NIR and MIR PL relations for LMC Cepheids corrected for reddening \label{tab7}
}

\tablehead{
\colhead{Wesenheit definition}&
\colhead{a  $\pm \sigma_a$}&
\colhead{b   $\pm \sigma_b$}&
\colhead{Num}&
\colhead{$\sigma$}
}
\startdata

\multicolumn{5}{c}{FU CEPHEIDS} \\
\\

$V$\tablenotemark{a}  & 17.172 $\pm$ 0.001 & -2.807 $\pm$ 0.001 &         1526 &  0.08\\
$I$\tablenotemark{a}  & 16.674 $\pm$ 0.001 & -3.017 $\pm$ 0.001 &         1520 &  0.08\\
$J$\tablenotemark{a}  & 16.256 $\pm$ 0.001 & -3.068 $\pm$ 0.002 &         1516 &  0.09\\
$H$\tablenotemark{a}  & 16.102 $\pm$ 0.002 & -3.257 $\pm$ 0.003 &         1514 &  0.08\\
$K_{\rm{S}}$\tablenotemark{a}   & 16.053 $\pm$ 0.002 & -3.261 $\pm$ 0.003 &         1518 &  0.09\\
$W1$\tablenotemark{a}  & 15.864 $\pm$ 0.006 & -3.194 $\pm$ 0.007 &         1493 &  0.17\\
$V$\tablenotemark{b}  & 17.272 $\pm$ 0.002 & -2.722 $\pm$ 0.003 &         1118 &  0.18\\
$I$\tablenotemark{b}  & 16.740 $\pm$ 0.002 & -2.963 $\pm$ 0.003 &         1112 &  0.13\\
$J$\tablenotemark{b}  & 16.314 $\pm$ 0.001 & -3.088 $\pm$ 0.002 &         1121 &  0.11\\
$H$ \tablenotemark{b} & 16.111 $\pm$ 0.002 & -3.227 $\pm$ 0.003 &         1129 &  0.09\\
$K_{\rm{S}}$\tablenotemark{b}   & 16.069 $\pm$ 0.002 & -3.245 $\pm$ 0.003 &         1134 &  0.09\\
$V$\tablenotemark{c}    & 17.435 $\pm$ 0.002 & -2.672 $\pm$ 0.002 &         1523 &  0.23\\
$I$\tablenotemark{c}    & 16.820 $\pm$ 0.002 & -2.909 $\pm$ 0.002 &         1516 &  0.15\\
$J$\tablenotemark{c}    & 16.341 $\pm$ 0.001 & -3.001 $\pm$ 0.002 &         1520 &  0.12\\
$H$\tablenotemark{c}    & 16.163 $\pm$ 0.002 & -3.244 $\pm$ 0.003 &         1520 &  0.10\\
$K_{\rm{S}}$\tablenotemark{c}     & 16.097 $\pm$ 0.002 & -3.249 $\pm$ 0.003 &         1526 &  0.10\\
$w1$\tablenotemark{c}    & 15.853 $\pm$ 0.006 & -3.156 $\pm$ 0.007 &         1498 &  0.18\\

  \hline
		    \\
\multicolumn{5}{c}{FO CEPHEIDS} \\
\\
$V$\tablenotemark{a}   & 16.789 $\pm$ 0.001 & -3.080 $\pm$ 0.003 &         1056 &  0.09\\
$I$\tablenotemark{a}   & 16.234 $\pm$ 0.001 & -3.201 $\pm$ 0.003 &         1063 &  0.09\\
$J$\tablenotemark{a}   & 15.842 $\pm$ 0.002 & -3.312 $\pm$ 0.005 &         1067 &  0.11\\
$H$\tablenotemark{a}   & 15.512 $\pm$ 0.006 & -3.265 $\pm$ 0.01 &         1059 &  0.11\\
$K_{\rm{S}}$\tablenotemark{a}    & 15.535 $\pm$ 0.003 & -3.330 $\pm$ 0.008 &         1035 &  0.12\\
$W1$\tablenotemark{a}   & 15.302 $\pm$ 0.001 & -3.210 $\pm$ 0.003 &         1038 &  0.19\\
$V$\tablenotemark{b}   & 16.860 $\pm$ 0.001 & -3.299 $\pm$ 0.004 &          795 &  0.19\\
$I$\tablenotemark{b}   & 16.285 $\pm$ 0.001 & -3.334 $\pm$ 0.004 &          797 &  0.14\\
$J$\tablenotemark{b}   & 15.878 $\pm$ 0.002 & -3.389 $\pm$ 0.005 &          797 &  0.12\\
$H$\tablenotemark{b}   & 15.626 $\pm$ 0.005 & -3.455 $\pm$ 0.01 &          790 &  0.10\\
$K_{\rm{S}}$\tablenotemark{b}    & 15.587 $\pm$ 0.003 & -3.455 $\pm$ 0.007 &          788 &  0.10\\
$V$\tablenotemark{c}  & 16.963 $\pm$ 0.001 & -3.141 $\pm$ 0.003 &          930 &  0.23\\
$I$\tablenotemark{c}  & 16.344 $\pm$ 0.001 & -3.240 $\pm$ 0.003 &          931 &  0.16\\
$J$\tablenotemark{c}  & 15.914 $\pm$ 0.002 & -3.334 $\pm$ 0.005 &          930 &  0.13\\
$H$\tablenotemark{c}  & 15.533 $\pm$ 0.006 & -3.246 $\pm$ 0.01 &          928 &  0.12\\
$K_{\rm{S}}$\tablenotemark{c}   & 15.567 $\pm$ 0.004 & -3.336 $\pm$ 0.009 &          910 &  0.13\\
$w1$\tablenotemark{c}  & 15.383 $\pm$ 0.001 & -3.322 $\pm$ 0.004 &          897 &  0.19\\

\enddata

\tablenotetext{a}{Reddening correction performed by adopting our estimates of $E(B-V)$}
\tablenotetext{b}{Reddening correction performed by adopting $E(V-I)$ taken from H11}
\tablenotetext{c}{No reddening correction performed}

\end{deluxetable}

 \clearpage  

\begin{deluxetable}{lcclllllllllllll}
\tabletypesize{\scriptsize}
\tablecaption{Example line of the published online catalog (col. 1-12) \label{tab9} }
\tablehead{
 \colhead{ID\tablenotemark{a}} &
  \colhead{Mode} &
  \colhead{Sample\tablenotemark{b}} &
  \colhead{Period} &
  \colhead{$I$\tablenotemark{c}} &
  \colhead{$V$\tablenotemark{c}} &
  \colhead{$J$\tablenotemark{c}} &
  \colhead{$\sigma_J$\tablenotemark{d}} &
  \colhead{$H$\tablenotemark{c}} &
  \colhead{$\sigma_H$\tablenotemark{d}} &
  \colhead{$Ks$\tablenotemark{c}} &
  \colhead{$\sigma_{Ks}$\tablenotemark{d}} \\
    }
\startdata
&&&&&&&&&&&\\
HV6098 & FU & P04 & 24.238 & 12.27 & 12.95 & 11.717 & 0.017 & 11.395 & 0.016 & 11.303 & 0.014 \\
&&&&&&&&&&&\\
\enddata    
\tablenotetext{a}{Name of the star: the the Harvard Variable catalog ID for the first ten Cepheids, and the OGLE ID for all the others in the OGLE catalog.}
\tablenotetext{b}{The sub-sample to which the Cepheid belongs, as defined in Table~1.}
\tablenotetext{c}{The mean intensity transformed into magnitude.}
\tablenotetext{d}{Error on the mean intensity transformed into magnitude.}
\end{deluxetable} 

\begin{deluxetable}{llllccccccc}
\tabletypesize{\scriptsize}
\tablenum{9}
\tablecaption{(Cont.) Example line of the published online catalog (col. 13-23)}
\tablehead{
  \colhead{$\alpha$} &
    \colhead{$\delta$} &
 \colhead{$w1$} &
  \colhead{$\sigma_{w1}$} &
  \colhead{$w1_{flag}$\tablenotemark{e}} &
  \colhead{$w1_{Nobs}$\tablenotemark{f}} &
  \colhead{DM\tablenotemark{g}} &
 \colhead{$\sigma_{DM}$} &
  \colhead{$E(B-V)$\tablenotemark{h}} &
  \colhead{$\sigma_{E}$} &
  \colhead{$\chi^2$\tablenotemark{j}} \\
}
\startdata
&&&&&&&&&&\\
74.437668  &-65.708359&11.247 & 0.035 & Y & 95 & 18.01 & 0.03 & 0.059 & 0.029 & 48\\
&&&&&&&&&&\\
\enddata    

\tablenotetext{e}{A flag that specifies if the $w1$-band mean magnitude was obtained
by performing the Fourier-fit (Flag = Y) or by adopting the error-weighted mean of the observed magnitudes (Flag = N).}
\tablenotetext{f}{Number of epochs available in the $w1$ band.}
\tablenotetext{g}{Distance modulus (in magnitude) obtained by adopting the reddening-fit method.}
\tablenotetext{h}{Color excess (in magnitude) obtained by adopting the reddening-fit method.}
\tablenotetext{j}{The $\chi^2$ of the reddening-law fit.}

\end{deluxetable} 

\clearpage

\clearpage

\end{document}